\documentclass[aps,prl,twocolumn,preprintnumbers,amsmath,amssymb,showkeys,floatfix,nofootinbib,superscriptaddress, reprint]{revtex4-1}

\usepackage{amsmath}
\usepackage{amsfonts}
\usepackage{amssymb}
\usepackage{graphicx}
\usepackage{color}
\usepackage{bm}
\usepackage{hyperref}
\usepackage{diagbox}

\begin{document}

%
%
% -------------------------------- our notations --------------------------- %
%

\def\ada#1{\textcolor{blue}{#1}}
\def\jonas#1{\textcolor{red}{#1}}

\def\ket#1{ $ \left\vert  #1   \right\rangle $}
\def\ketm#1{  \left\vert  #1   \right\rangle   }
\def\bra#1{ $ \left\langle  #1   \right\vert $ }
\def\bram#1{  \left\langle  #1   \right\vert   }
\def\spr#1#2{ $ \left\langle #1 \left\vert \right. #2 \right\rangle $ }
\def\sprm#1#2{  \left\langle #1 \left\vert \right. #2 \right\rangle   }
\def\me#1#2#3{ $ \left\langle #1 \left\vert  #2 \right\vert #3 \right\rangle $}
\def\mem#1#2#3{  \left\langle #1 \left\vert  #2 \right\vert #3 \right\rangle   }
\def\redme#1#2#3{ $ \left\langle #1 \left\Vert
                  #2 \right\Vert #3 \right\rangle $ }
\def\redmem#1#2#3{  \left\langle #1 \left\Vert
                  #2 \right\Vert #3 \right\rangle   }
\def\threej#1#2#3#4#5#6{ $ \left( \matrix{ #1 & #2 & #3  \cr
                                           #4 & #5 & #6  } \right) $ }
\def\threejm#1#2#3#4#5#6{  \left( \matrix{ #1 & #2 & #3  \cr
                                           #4 & #5 & #6  } \right)   }
\def\sixj#1#2#3#4#5#6{ $ \left\{ \matrix{ #1 & #2 & #3  \cr
                                          #4 & #5 & #6  } \right\} $ }
\def\sixjm#1#2#3#4#5#6{  \left\{ \matrix{ #1 & #2 & #3  \cr
                                          #4 & #5 & #6  } \right\} }

\def\ninejm#1#2#3#4#5#6#7#8#9{  \left\{ \matrix{ #1 & #2 & #3  \cr
                                                 #4 & #5 & #6  \cr
                         #7 & #8 & #9  } \right\}   }
%
%
% ---------------------------- end of our notations --------------------------- %
%
%

%
% -----------------------------------------    Title of paper ---------------------------------------
%

\title{Quantum effects on plasma screening for thermonuclear reactions in laser-generated plasmas}

%
% ------------------------------------------   List of authors --------------------------------------
%

\author{David \surname{Elsing}}
\altaffiliation[Present address: ]{Institute of Nanotechnology, KIT Campus North, Hermann-von-Helmholtz-Platz 1, 76344 Eggenstein-Leopoldshafen, Germany}
\affiliation{Max-Planck-Institut f\"ur Kernphysik, Saupfercheckweg 1, D-69117 Heidelberg, Germany}

\author{Adriana \surname{P\'alffy}}
\affiliation{Max-Planck-Institut f\"ur Kernphysik, Saupfercheckweg 1, D-69117 Heidelberg, Germany}
\affiliation{Department of Physics, Friedrich-Alexander-Universit\"at  Erlangen-N\"urnberg,  D-91058 Erlangen, Germany}

\author{Yuanbin \surname{Wu}}
\email[Corresponding author.]{yuanbin.wu@mpi-hd.mpg.de}
\affiliation{Max-Planck-Institut f\"ur Kernphysik, Saupfercheckweg 1, D-69117 Heidelberg, Germany}

\date{\today}

%
%
%
% ---------------------------------------------------- Abstract ---------------------------------------------
%
%
%
%
\begin{abstract}

A quantum plasma screening model based on the density matrix formalism is used to investigate theoretically the thermonuclear reactions $^{13}$C($\alpha$, $n$)$^{16}$O and $^2$H($d$, $n$)$^3$He in laser-generated plasmas over a large range of densities and temperatures. For cold and dense (solid-state density) plasmas, our results show that quantum effects can enhance the plasma screening for thermonuclear reactions up to one order of magnitude compared to the classical case. This result can have impact on nuclear astrophysics predictions, and also may play a role for fusion energy gain prospects. Our simulations allow us to identify the laser-generated plasma experimental setting in which the quantum effects on plasma screening could be  confirmed at existing high-intensity laser facilities.

\end{abstract}

\maketitle

%%%----------------------- introduction -------------------------

{\it Introduction.} In plasmas, long-range electric fields are screened by the dynamic flow of moving particles. This charge screening enhances the nuclear reaction cross sections by reducing the Coulomb barrier that reacting ions must overcome \cite{AdelbergerRMP2011, AtzeniBook2004}. Plasma screening is 
 a crucial aspect for thermonuclear reactions in astrophysical plasmas such as  star cores where nucleosynthesis occurs \cite{AdelbergerRMP2011}, but also for industrial fusion energy gain \cite{AtzeniBook2004}, for instance inertial confinement fusion \cite{Labaune2013,HurricaneNature2014, Labaune2015,OlsonPRL2016, CerjanJPG2018}   that may provide future sources of alternative energy. Theoretical plasma screening models focus mostly on classical approaches \cite{Salpeter1954AJP, MitlerAPJ1977, Dzitko1995APJ, Salpeter1969APJ, DewittAPJ1973, Graboske1973APJ}, while quantum plasma models only address the weak screening regime and have shown good agreement with the classical weak screening \cite{Gruzinov1998APJ, WiletsAPJ2000, Bahcall2002AA, Chitanvis2007APJ}.

  Hints on the important role of quantum effects come from the atomic counterpart of plasmas screening,  the plasma-induced ionization potential depression (IPD).  Direct measurements \cite{CiricostaPRL2012}  have been shown to conflict the extensively used Stewart-Pyatt IPD model \cite{StewartAPJ1966} which interpolates between the limits of the Debye-H\"{u}ckel theory  \cite{DebyePZ1923, GriemBook1964} and the ion-sphere model \cite{Salpeter1954AJP}. The debate is still ongoing  \cite{HoartyPRL2013, FletcherPRL2014, CiricostaNC2016, KrausPPCF2019} and has shed light on the role of  quantum effects \cite{VinkoNC2014,JinPRE2021,ZengAA2020}.  On the front of plasma screening in thermonuclear reactions, the lack of experimental evidence could not resolve several controversies \cite{Tsytovich2000,AdelbergerRMP2011, CerjanJPG2018, NegoitaRRP2016},  despite numerous theoretical studies  \cite{Salpeter1954AJP, MitlerAPJ1977, Dzitko1995APJ, Salpeter1969APJ, DewittAPJ1973, Graboske1973APJ, Gruzinov1998APJ, Keller1953APJ, BrownRMP1997, GruzinovAVApJ1998b, WiletsAPJ2000, Bahcall2002AA, KravchukPRC2014, Chitanvis2007APJ, Shaviv1996APJ, Shaviv2000APJ, Mao2009APJ, KushnirMNRAS2019, ClerouinPoP2019}.  Fortunately, the development of laser technology in the past decades promises the appropriate conditions for conclusive experiments in the lab. Both    X-ray Free Electron Lasers (XFELs) \cite{europeanXFEL-web, LCLS-web, SwissFEL-web, SACLA-web} and ultra-strong optical lasers \cite{DansonHPLSE2014, DansonHPLSE2019, ELI-web, LULI-web, Vulcan-web, NIF-web} open so-far unavailable parameter regimes  for the study of nuclear physics in laser-generated plasmas \cite{Labaune2013,CerjanJPG2018, NegoitaRRP2016, CaseyNP2017, GunstPRL2014, GunstPOP2015}. In particular, theoretical predictions show that experiments at petawatt optical lasers should allow tests of the 
widely used Salpeter weak screening model for thermonuclear reactions \cite{Wu2017APJ, WuPOP2020}.

In this Letter we investigate the role of  quantum effects for screening in laser-generated plasmas in the intermediate screening regime of low temperature and high density. This  regime has  became available experimentally at newly commissioned laser facilities and is  relevant for the evolution of low mass stars, brown dwarfs, and pre-main-sequence stars as well as the lithium depletion problem \cite{PaxtonAPJS2011, HidalgoAPJ2018, ForbesAPJ2019, MoussaEPL2017, AguileraAPJ2016, SomersAPJ2016, TognelliMNRAS2015, YoungAPJ2003, ChabrierARAA2000, ChabrierAA1997, LagardeAA2012, AmardAA2019}. We  employ the density matrix formalism (DMF) derived in quantum statistical mechanics \cite{Gruzinov1998APJ}
to include quantum effects in plasma screening and compare our results with classical predictions over a large range of densities and temperatures accessible in laser-generated plasmas. Surprisingly, in the intermediate screening regime the DMF quantum plasma model predicts up to one order of magnitude higher screening factors than the classical plasma models. This enhancement is sufficiently large to be observed experimentally in laser-generated plasma experiments. We investigate three realistic experimental settings at existing petawatt and x-ray laser facilities and determine nuclear reaction rates for currently accessible experimental parameters. Based on our predictions, we identify the most promising experimental scenario and put forward an experimental test of quantum effects in the intermediate screening regime.

As case studies we choose  two thermonuclear reactions: (i) $^{13}$C($\alpha$, $n$)$^{16}$O    which is one of the important helium burning processes as well as one of the main neutron sources for the s-process  \cite{DeLooreBook1992, GallinoAPJ1998, HeilPRC2008, TrippellaAPJ2014, AliottaEPJA2016, CristalloAPJ2018}, and (ii) $^2$H($d$, $n$)$^3$He,  one of the key reactions in the study of inertial confinement fusion \cite{HurricaneNature2014, OlsonPRL2016, CerjanJPG2018}. 
 The quantum screening results are compared with  the Salpeter weak screening \cite{Salpeter1954AJP} and the Mitler formula \cite{MitlerAPJ1977, Dzitko1995APJ}, both based on classical plasma models. In addition, for the $^{13}$C($\alpha$, $n$)$^{16}$O reaction in a cold and dense plasma  case we also compare our results with interpolation formulae for the Salpeter and Von Horn intermediate screening (SVH) \cite{Salpeter1969APJ} and the GDGC classical plasma model by Graboske {\it{et al.}} \cite{Graboske1973APJ}. These models use numerical interpolation of classical model results for the intermediate screening regime. Our numerical results show that all models display good agreement for low plasma densities and high temperatures, confirming previous results \cite{Gruzinov1998APJ, WiletsAPJ2000, Bahcall2002AA, Chitanvis2007APJ}. However, for low temperature and high density, the DMF quantum plasma model presents a substantial enhancement of the plasma screening. This  holds true also for the comparison with results of the Mitler formula \cite{MitlerAPJ1977, Dzitko1995APJ} which is based on the Stewart-Pyatt IPD model and should be valid for the full range of plasma parameters.

 %%%----------------------- theory -------------------------

{\it Theory.} We consider the fusion reaction of two positively charged nuclei with charge numbers $Z_1$ and $Z_2$. Due to  screening, the nuclear reaction rate in plasmas can be enhanced by a factor $g_{\rm{scr}}$ \cite{AtzeniBook2004} defined as $<\!\! \sigma v \!\!>_{\rm{scr}} = g_{\rm{scr}} <\!\! \sigma v \!\!>$, where $<\!\! \sigma v \!\!>$ is the the averaged reactivity neglecting screening, $\sigma$ is the nuclear reaction cross section and $v$ the particle relative velocity, respectively. $<\!\! \sigma v \!\!>$ is given by the averaging of $\sigma v$ over the reactant velocity distribution. In weakly coupled plasmas, i.e., plasmas in which the Coulomb interaction energy between the nucleus and the nearest few electrons and nuclei is small compared to the thermal energy, the classical Salpeter model  \cite{Salpeter1954AJP, Gruzinov1998APJ, Wu2017APJ}  gives the plasma screening enhancement factor (in atomic units with the Boltzmann constant $k_B=1$)  $g_{\rm{scr}} = \exp{\left[ Z_1 Z_2/(\lambda_D T) \right]}$,
where $\lambda_D$ is the Debye length, and $T$ is the plasma temperature. This holds for low-density and high-temperature plasmas. For dense plasmas, Salpeter applied the ion-sphere approximation to the strong screening \cite{Salpeter1954AJP}. The intermediate regime can be described by numerical interpolations in the SVH approach \cite{Salpeter1969APJ}. 

Starting from the Stewart-Pyatt IPD model \cite{StewartAPJ1966}, Mitler considered the charge density to be constant for small distances close to the nucleus (ion-sphere model), while applying the Debye-H\"{u}ckel theory at large distances \cite{MitlerAPJ1977, Dzitko1995APJ}. This lead to an expression valid over the entire range of plasma parameters for the   screening enhancement factor \cite{MitlerAPJ1977, Dzitko1995APJ}
\begin{equation} \label{eq:mitlers}
  g_{\rm{scr}} = \exp{\left[  8\pi^2 n_e^2 \lambda_D^5 \left|\Delta G\right|/(5T)  \right]},
\end{equation}
where $n_e$ is the electron density and 
\begin{equation}
  \Delta G = (\zeta_{Z_1} + \zeta_{Z_2} + 1)^{\frac{5}{3}} - (\zeta_{Z_1}+ 1)^{\frac{5}{3}} - (\zeta_{Z_2} + 1)^{\frac{5}{3}} + 1,
\end{equation}
with $\zeta_Z = 3Z/(4\pi n_e \lambda_D^3)$. 

We now turn to  the DMF quantum plasma model  introduced in Ref.~\cite{Gruzinov1998APJ}. In this model, the electron density is derived via the density matrix equation known in  quantum statistical mechanics \cite{Gruzinov1998APJ, FeynmanBook1972},
\begin{equation} \label{eq:dmfden}
\frac{\partial \rho({\bf{r}}', {\bf{r}}; \beta)}{\partial{\beta}} = \left[ \nabla_{{\bf{r}}'}^2 /2+ \Phi({\bf{r}}') \right] \rho({\bf{r}}', {\bf{r}}; \beta)
\end{equation}
with the initial condition $\rho({\bf{r}}', {\bf{r}}; \beta = 0) = \delta^{(3)} ({\bf{r}}' - {\bf{r}})$, where $\beta = 1/T$ and $\Phi$ is the potential around the nuclear charge. The electron density should be normalised by the solution of Eq.~\eqref{eq:dmfden} for nuclear charge $Z = 0$, $\rho_0(\beta) = (2\pi \beta)^{-3/2}$ \cite{Gruzinov1998APJ}. Using this normalisation, the  total electron density becomes $\rho_e ({\bf{r}}) = n_e (2\pi \beta)^{3/2} \rho({\bf{r}}, {\bf{r}}; \beta)$. The potential $\Phi$ surrounding a nuclear charge $Z$ is then described by the modified Poisson-Boltzmann equation \cite{Gruzinov1998APJ}
\begin{equation} \label{eq:dmfpot}
  \nabla^2 \delta\Phi = -4\pi \left[ n_b \sum_i \frac{Z_i X_i}{A_i} \exp{(-\beta Z_i \Phi)} -  \rho_e \right],
\end{equation}
where $\delta\Phi = \Phi -Z/r$, $n_b$ is the baryon density, and $X_i$ and $A_i$ are the mass fraction and the mass number of $i$-species ion, respectively. Equations \eqref{eq:dmfden} and \eqref{eq:dmfpot} can be solved self-consistently.
We note that the spatial dependence of the screening potential has also been considered in the WKB approximation in Refs.~\cite{ItohPTPS1981,ItohApJ1977,ItohApJ1979}, however, for the strong screening regime. In our case, 
with the solution of Eqs. \eqref{eq:dmfden} and \eqref{eq:dmfpot} for the density distribution and potential, the plasma screening enhancement factor can be obtained  as
 $ g_{\rm{scr}} = \exp{\left[ -\beta \left( F_{Z_1 + Z_2} - F_{Z_1} - F_{Z_2} \right)  \right]}$
where, $F_Z$ stands for the free energy obtained  in terms of the electrostatic energy $U_Z$ via $F_Z = \frac{1}{\beta} \int_0^{\beta} U_Z(\tau) d\tau$ \cite{Gruzinov1998APJ}. 

%%%%%%%%%%%%%%%%%%%%%%%%%%%%%%%%%%%%%%%%%%%%%%%%%%%

{\it Numerical approach.}    The numerical approach for solar plasma parameters described in Ref.~\cite{Gruzinov1998APJ}  fails to provide accurate electron densities for small temperatures and high densities. For this parameter regime, the numerical integration of Eq.~\eqref{eq:dmfden}  becomes  cumbersome for two reasons: (1) The initial value is a Dirac $\delta$ function, which is approximated by a Gaussian with an appropriate width, requiring a small grid spacing near ${\bf{r}}' = \bf{r}$; (2) The potential $\Phi({\bf{r}}')$ diverges at ${\bf{r}}' = 0$ and has to be approximated numerically by using a regularization procedure \cite{Gruzinov1998APJ}.   The increasing error stemming from the $\beta$ integration in Eq.~\eqref{eq:dmfden} becomes quickly noticeable when going to smaller temperatures.

Eq.~\eqref{eq:dmfden} has cylindrical symmetry and is solved on a non-linear 
two-dimensional grid in cylindrical coordinates $R$ and $z$.  Since the grid spacing is very small at the points where  $\partial\rho/\partial \beta$ is very large, we combine an implicit Crank-Nicolson step size \cite{CrankMPCPS1947} with the Runge-Kutta-Fehlberg method \cite{HairerBook1993}. 
 For large ${\bf{r}}'$, $\Phi$ is approximated by the weak screening potential of the Mitler model, while the density matrix is  set to zero for the finite differences at the boundaries of the numerically considered ${\bf{r}}'$ region. Between the iterations in solving Eqs. \eqref{eq:dmfden} and \eqref{eq:dmfpot}, both the density and the potential are interpolated to the new grid by cubic spline interpolation.  With the numerical procedure described above, free energies $F_Z$ obtained agree with the ones in Ref.~\cite{Gruzinov1998APJ} on the level of a few percent.

 % ............ continue here! ................................

%%%----------------------- results and discussions -------------------------

\begin{figure}[h!]
\centering
\includegraphics[width=1.0\linewidth]{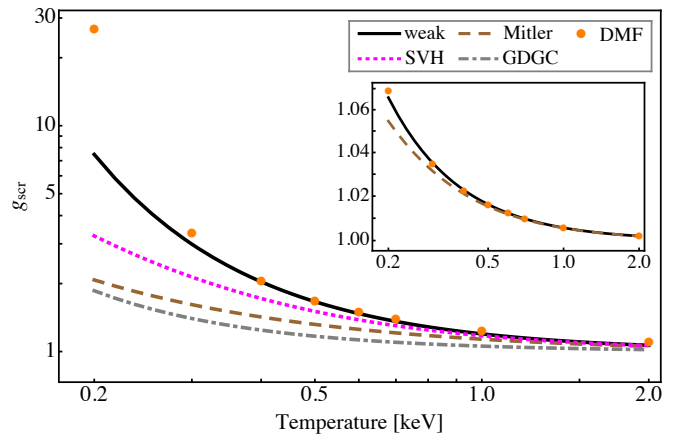}%
\caption{Plasma screening enhancement factor $g_{\rm{scr}}$ for the  $^{13}$C($\alpha$, $n$)$^{16}$O reaction as a  function of the He plasma temperature for 
 weak screening \cite{Salpeter1954AJP} (black solid curve), Mitler formula \cite{MitlerAPJ1977}  (brown dashed curve), SVH interpolation  
\cite{Salpeter1969APJ} (magenta dotted curve), GDGC interpolation \cite{Graboske1973APJ} (grey dash-dotted curve), and the DMF model  (orange filled circle). We consider the plasma density 
 $10^{24}$ cm$^{-3}$ ($10^{21}$ cm$^{-3}$ for inset).   
\label{fig:chetemp}}
\end{figure}

{\it Numerical results.} We calculate the plasma screening enhancement factor $g_{\rm{scr}}$ for the  $^{13}$C($\alpha$, $n$)$^{16}$O reaction occurring in a Helium plasma. 
Figure~\ref{fig:chetemp} presents the calculated screening enhancement factor compared to classical model results as a function of  plasma temperature  for He number densities $10^{24}$ cm$^{-3}$ and $10^{21}$ cm$^{-3}$ (in the inset).  
  The results  show that for lower density, the DMF model predicts similar enhancement factors as the weak (Salpeter) screening model, confirming previous predictions ~\cite{Gruzinov1998APJ}. At the same time, the Mitler formula predicts sightly smaller enhancement factors than the weak screening at low temperatures, but the overall differences between  models remain small. Thus, no significant quantum effects are expected 
to affect experiments based on the $^{13}$C($\alpha$, $n$)$^{16}$O reaction aimed at determining $g_{\rm{scr}}$ in the weak screening regime \cite{Wu2017APJ}.

\begin{figure}[h!]
\centering
\includegraphics[width=1.0\linewidth]{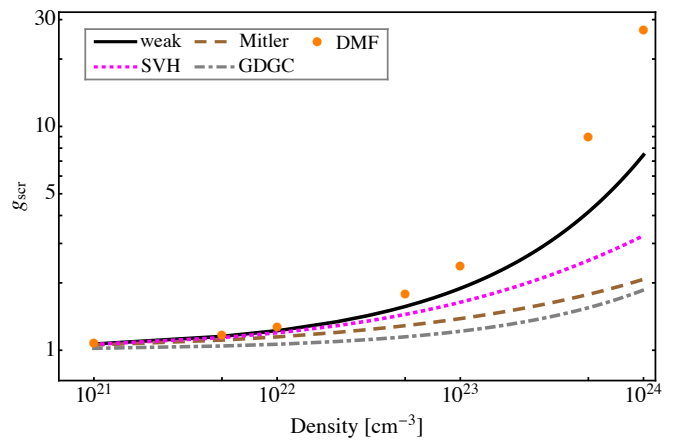}%
\caption{Plasma screening enhancement factor $g_{\rm{scr}}$ for the $^{13}$C($\alpha$, $n$)$^{16}$O reaction as function of helium plasma  density at  temperature $200$ eV for the same models as in Fig.~\ref{fig:chetemp}.
\label{fig:cheden}}
\end{figure}

    For the high density case, the five plasma screening models show good overall agreement only for high temperatures and spread out significantly for temperatures of few hundreds eV. Here,  DMF predicts the highest plasma screening enhancement factor among the considered models. The Mitler, SVH and GDGC predictions lie below the Salpeter weak screening, which in turn remains one order of magnitude lower than the DMF results at 200 eV temperature. Figure 
 \ref{fig:cheden} shows the density dependence of the plasma screening factor at this temperature and reveals that 
 the disagreement between models becomes increasingly visible starting with densities of few times $10^{22}$ cm$^{-3}$.  We note that the Mitler formula is  expected to be valid for the full range of plasma parameters, while the intermediate screening SVH formula is expected to have an error on the level of $30 \%$ for the cases of  screening enhancement factor around the value of $e$, the Euler number \cite{Salpeter1969APJ}. In addition, the GDGC intermediate screening formula obtained assuming completely degenerate electrons \cite{DewittAPJ1973, Graboske1973APJ}, would underestimate the screening effect \cite{Gruzinov1998APJ} when applied to our case which is only weakly degenerate.

 From Figs. \ref{fig:chetemp} and \ref{fig:cheden}, we can conclude that the quantum plasma effects become significant in the intermediate screening regime where the Salpeter weak screening also deviates significantly from the other classical plasma models. For both classical and quantum models, the low temperature and high density regime highlights the Coulomb effects  in the  immediate vicinity of the reacting nuclei. It is this physical region where the model assumptions for the Salpeter weak screening and the Mitler formula differ most, and where the quantum effects originate. Thus, low temperature (approx. 200 eV) and increasing density $>10^{23}$~cm$^{-3}$ are the physical conditions under which the  quantum plasma effects become important.
   This in turn indicates that  laser experiments based on gas jet targets cannot distinguish between screening models, as the highest density of gas jets achieved at present is approx.~$10^{21}$ cm$^{-3}$ \cite{SchmidRSI2012, SyllaRSI2012}. Solid-state density plasma targets should be used instead.

%%%%%%%%%%%%%%%%%%%%%%%%%%%%%%%%%%%%%

%%%%%%%%%%%%%%%%%%%%%%%%%%%%%%%%%%%%%% 

%%%%%%%%%%%%%%%%%%%%%%%%%%%%%%%%%%%%%
%%%%%%%%%%%%%%%%%%%%%%%%%%%%%%%%%%%%%% 

%%%%%%%%%%%%%%%%%%%%%%%%%%%%%%
\renewcommand{\arraystretch}{1.5}
\begin{table}
  \centering
  \footnotesize
  \begin{tabular}{lcccc}
  \hline\hline
  & & & & \tabularnewline[-0.4cm]
 [$n_i$, T] & [$10^{24}$, $0.2$] & [$10^{24}$, $0.4$] & [$10^{21}$, $0.2$] & [$10^{21}$, $0.4$] \tabularnewline

  & & & &  \tabularnewline[-0.4cm] \hline
  & & & &  \tabularnewline[-0.4cm]
weak & $1.10$ & $1.035$ & $1.0031$ & $1.0011$ \tabularnewline
  
Mitler & $1.079$ & $1.031$ & $1.0030$ & $1.0011$ \tabularnewline

DMF & $1.12$ & $1.050$ & $1.0030$ & $1.0011$ \tabularnewline

  & & & & \tabularnewline[-0.4cm] \hline\hline
  \end{tabular}
  \caption{Plasma screening enhancement factors $g_{\rm{scr}}$ for the reaction $^2$H($d$, $n$)$^3$He. A deuterium plasma  with ion density $n_i$ in units of cm$^{-3}$ is assumed. The plasma temperature $T$ is given in units of keV. Salpeter weak screening \cite{Salpeter1954AJP}, Mitler formula \cite{MitlerAPJ1977}, and the present DMF model are considered.}
  \label{table:ddscr}
\end{table}
\renewcommand{\arraystretch}{1}
%%%%%%%%%%%%%%%%%%%%%%%%%%%%%%%

We now turn to the second investigated thermonuclear reaction  $^2$H($d$, $n$)$^3$He occurring in a deuterium plasma.  $^2$H($d$, $n$)$^3$He  is one of the key reactions in the study of the inertial confinement fusion \cite{HurricaneNature2014, OlsonPRL2016, CerjanJPG2018}. Calculated plasma screening enhancement factors $g_{\rm{scr}}$ using the Salpeter week screening, the Mitler formula and the DMF model are presented in Table~\ref{table:ddscr} for selected plasma density and temperature values. We observe also in this case the same general trend. While the three models agree very well for $k_BT= 400$~eV and deuterium ion density $n_i=10^{21}$ cm$^{-3}$, discrepancies occur going to larger densities and lower temperatures. The predicted values for $n_i=10^{21}$ cm$^{-3}$ show that  also for the $^2$H($d$, $n$)$^3$He reaction, the screening effect would be negligible in gas-jet experiments which are bound to low ion density values, as already implied in Ref.~\cite{WuPOP2020}. For higher plasma densities, the DMF model provides the largest screening factor, while the Mitler formula gives the smallest one. However, the disagreements are on the order of only 2\% even for a plasma temperature of 200 eV and plasma density  $10^{24}$ cm$^{-3}$. The quantum plasma effects are weaker in the case of $^2$H($d$, $n$)$^3$He compared with $^{13}$C($\alpha$, $n$)$^{16}$O. This can be explained by the lower charges of the nuclear reactants and therefore the weaker Coulomb fields in the immediate vicinity of the nuclei. The small quantum effects are correlated with a better agreement between the Salpeter and Mitler screening factors, showing that for the $^2$H($d$, $n$)$^3$He reaction, the intermediate screening regime would require even higher density and lower temperature.

%------------ experimental possibilities ................................

{\it Possible experimental verification.} Recent works have shown that solid-state density targets can be heated to temperatures starting from a few hundred eV up to a few keV either via isochoric heating \cite{Saemann1999PRL,Audebert2002PRL,Sentoku2007POP,Wu2018PRL} or by using XFELs \cite{LeeJOSAB2003,VinkoNature2012}. Thus, the intermediate regime screening conditions should be accessible experimentally. 
In the following we address three realistic scenarios that promise to shed light on the screening mechanism and the role of quantum effects. In all cases, determining the screening factor requires three successive experiments: the thermonuclear reaction occurring in the setup (1) in the absence of a plasma target, (2) with a low-density and (3) with a high-density plasma target, the latter two at the same temperature. In parallel, plasma diagnostics on  density, temperature, and energy distributions will be necessary in order to allow for data reconstruction. 
By comparison between the three experimental outcomes and with theory, the value of the plasma screening factor could be deduced \cite{Wu2017APJ}.

{\it Setup I.}  At mega-Joule laser facilities such as the National Ignition Facility (NIF) in the US \cite{NIF-web}, thermonuclear reactions have been observed in solid-state or even higher density plasmas with temperatures from a few hundred eV to a few keV \cite{HurricaneNature2014, OlsonPRL2016, CerjanJPG2018, CaseyNP2017, HayesNP2020}. However, the considered $^{13}$C($\alpha$, $n$)$^{16}$O reaction has the disadvantage that  experimental cross section data  is not available for few hundreds eV energies  \cite{XuNPA2013}, rendering unreliable extrapolations to a region of very small cross sections necessary.  The situation is more promising for the  reaction $^2$H($d$, $n$)$^3$He, for which cross sections have been well measured experimentally down to rather low energy \cite{XuNPA2013}. To give a numerical estimate, we follow the spherical plasma model in Ref.~\cite{WuPOP2020}. We assume a laser energy of $1$ MJ per pulse, deuterium ion density $10^{24}$ cm$^{-3}$,  plasma temperature is $200$ eV, and the interaction time  $1$ ns. The calculated $^2$H($d$, $n$)$^3$He event numbers per pulse are $4.01 \cdot 10^7$ s$^{-1}$ for the DMF model, $3.94 \cdot 10^7$ s$^{-1}$ for the weak screening, and $3.86 \cdot 10^7$ s$^{-1}$ for the Mitler model, respectively. Unfortunately, these values are too close to identify the underlying plasma screening mechanism.

{\it Setup II}. At the upcoming Nuclear Pillar of the Extreme Light Infrastructure facility (ELI-NP) \cite{ELI-web}, an experimental setup involving two laser beams that generate two colliding plasmas has been proposed \cite{NegoitaRRP2016}.  A laser pulse interacting with a first solid-state target should produce a rapidly streaming projectile plasma by means of target normal sheath acceleration (TNSA). This particle beam interacts with the (secondary) target plasma generated by the interaction of a second laser pulse on a gas jet target \cite{NegoitaRRP2016}. This setup has been already successfully applied in the context of aneutronic fusion reactions \cite{Labaune2013}. Theoretical predictions show that at ELI-NP, this experimental design will be suitable  for  testing weak screening models \cite{Wu2017APJ}. For the parameter regime investigated here, a modified setup  could be employed by changing the secondary gas jet target to a liquid drop target or a solid target \cite{DongNC2017} to test also the intermediary screening regime.  

For an ELI-NP beam for TNSA with 10 PW power, 25 fs pulse duration, 800 nm wavelength and intensity of $3\cdot 10^{19}$ W/cm$^2$, and a target thickness of 2 $\mu$m, our simulations show that an accelerated beam of C$^{5+}$ ions could be generated. We choose the parameters such that 
in the center-of-mass reference frame, the reaction energy is  shifted towards 1 MeV, in a region where the $^{13}$C($\alpha$, $n$)$^{16}$O cross sections have been measured experimentally with precision of a few percent \cite{HarissopulosPRC2005}  and astrophysical $S$-factor has a smooth energy dependence.  Using a plasma target of temperature 200 eV, thickness 10 $\mu$m and density $10^{24}$ cm$^{-3}$, the calculated reaction events per laser pulse are $4.4\cdot 10^5$ for the DMF, $1.2\cdot 10^5$ for weak screening and $3.4\cdot 10^4$ for the Mitler models, respectively. While the reaction numbers  lie apart by seizable factors, the experimental conditions at ELI-NP are likely to introduce temperature and density spatial and temporal gradients or non-thermal ion distributions, which are difficult to model theoretically or check experimentally.

{\it Setup III}. A strategy to mitigate the non-ideal plasma effects described above is to replace the second optical laser by an XFEL beam. Due to the high x-ray penetration power, the XFEL-generated plasma is expected to have more uniform conditions \cite{VinkoNature2012}.  The generation of cold and dense plasmas isochoricaly at XFEL has been demonstrated experimentally  \cite{LeeJOSAB2003,VinkoNature2012,CiricostaPRL2012}. We therefore address in the following a setup combining  optical and x-ray laser beams as the one at the Helmholtz International Beamline for Extreme Fields (HIBEF) \cite{HIBEF-web}. In this scenario, an optical laser would generate via TNSA the rapidly streaming projectile plasma, while the XFEL heats the secondary target to solid-state density  \cite{LeeJOSAB2003,VinkoNature2012,CiricostaPRL2012}. 
As suggested in Ref.~\cite{WuPRA2019}, in order to match XFELs in both power and repetition rate, mJ-class optical lasers are adopted.  We consider laser pulses with $100$ mJ energy, $100$ fs duration, and 800~nm wavelength. The intensity of the optical laser is assumed to be $2 \times 10^{19}$ W/cm$^2$. The optical laser interacts with a solid $^{13}$C target with a thickness of $2$ $\mu$m to generate a $^{13}$C ion beam. 
The ion beam interacts with a $^{4}$He plasma target generated by the XFEL with $^4$He ion density of $10^{24}$ cm$^{-3}$ and $200$ eV temperature. 
Also in this case we choose the parameters such that the center-of-mass reaction energy is shifted towards 1 MeV to avoid a complicated energy dependence of the astrophysical $S$-factor of the nuclear reaction.
Assuming a repetition rate of $10$ kHz for the optical laser and XFEL, we obtain the reaction rates $1.31 \cdot 10^7$ s$^{-1}$ for the DMF, $3.65 \cdot 10^6$ s$^{-1}$ for weak screening, and $1.02 \cdot 10^6$ s$^{-1}$ for the Mitler models, respectively. With the advantage of high repetition rates, more uniform plasma conditions, together with well measured nuclear reaction cross sections in relevant energies, and the large discrepancies of the reaction event rate between different screening models, this scenario could give a compelling evidence of the plasma screening model in the $^{13}$C($\alpha$, $n$)$^{16}$O reaction.

%%%----------------------- conclusion -------------------------

{\it Outlook.} The enhancement predicted by the quantum plasma screening model occurs in the intermediate screening regime, which could be rescaled to the astrophysical conditions relevant to the evolution of low mass stars, brown dwarfs, and pre-main-sequence stars as well as the lithium depletion problem \cite{PaxtonAPJS2011, HidalgoAPJ2018, ForbesAPJ2019, MoussaEPL2017, AguileraAPJ2016, SomersAPJ2016, TognelliMNRAS2015, YoungAPJ2003, ChabrierARAA2000, ChabrierAA1997, LagardeAA2012, AmardAA2019}. Such enhancement would lead to significantly impact in relevant realistic astrophysical scenarios, and hence would play important roles in the evolution of the above mentioned stars and the understanding of the lithium depletion problem. Furthermore, we may speculate that an experimental confirmation of quantum effects in plasma screening might shed light on the origin of the conflict between  direct IPD  measurements \cite{CiricostaPRL2012} and the extensively used Stewart-Pyatt IPD model \cite{StewartAPJ1966}. Once confirmed,  the quantum effect on the plasma screening enhancement may also play an important role for fusion energy gain prospects  \cite{Labaune2013,HurricaneNature2014, OlsonPRL2016, CerjanJPG2018,BerlinguetteNature2019}.

\begin{acknowledgments}
AP gratefully acknowledges support from the Heisenberg Program of the Deutsche  Forschungsgemeinschaft (DFG).

\end{acknowledgments}

%%%%%%%%%%%%%%%%%%%%%%%%%%%%%%%%%%%%%%%%%%%%%%%%%%%%%%%%%%

%%%%%%%%%%%%%%%%%%%%%%%%%%%%%%%%%%%%%%%%%%%%%%%%%%%%%%%%%

\bibliographystyle{apsrev-no-url-issn.bst}
\bibliography{refsscr}{}

\begin{thebibliography}{93}
\expandafter\ifx\csname natexlab\endcsname\relax\def\natexlab#1{#1}\fi
\expandafter\ifx\csname bibnamefont\endcsname\relax
  \def\bibnamefont#1{#1}\fi
\expandafter\ifx\csname bibfnamefont\endcsname\relax
  \def\bibfnamefont#1{#1}\fi
\expandafter\ifx\csname citenamefont\endcsname\relax
  \def\citenamefont#1{#1}\fi
\expandafter\ifx\csname url\endcsname\relax
  \def\url#1{\texttt{#1}}\fi
\expandafter\ifx\csname urlprefix\endcsname\relax\def\urlprefix{URL }\fi
\providecommand{\bibinfo}[2]{#2}
\providecommand{\eprint}[2][]{\url{#2}}

\bibitem[{\citenamefont{Adelberger et~al.}(2011)\citenamefont{Adelberger,
  Garc\'{\i}a, Robertson, Snover, Balantekin, Heeger, Ramsey-Musolf, Bemmerer,
  Junghans, Bertulani et~al.}}]{AdelbergerRMP2011}
\bibinfo{author}{\bibfnamefont{E.~G.} \bibnamefont{Adelberger}},
  \bibinfo{author}{\bibfnamefont{A.}~\bibnamefont{Garc\'{\i}a}},
  \bibinfo{author}{\bibfnamefont{R.~G.~H.} \bibnamefont{Robertson}},
  \bibinfo{author}{\bibfnamefont{K.~A.} \bibnamefont{Snover}},
  \bibinfo{author}{\bibfnamefont{A.~B.} \bibnamefont{Balantekin}},
  \bibinfo{author}{\bibfnamefont{K.}~\bibnamefont{Heeger}},
  \bibinfo{author}{\bibfnamefont{M.~J.} \bibnamefont{Ramsey-Musolf}},
  \bibinfo{author}{\bibfnamefont{D.}~\bibnamefont{Bemmerer}},
  \bibinfo{author}{\bibfnamefont{A.}~\bibnamefont{Junghans}},
  \bibinfo{author}{\bibfnamefont{C.~A.} \bibnamefont{Bertulani}},
  \bibnamefont{et~al.}, \bibinfo{journal}{Rev. Mod. Phys.}
  \textbf{\bibinfo{volume}{83}}, \bibinfo{pages}{195} (\bibinfo{year}{2011}).

\bibitem[{\citenamefont{Atzeni and Meyer-ter Vehn}(2004)}]{AtzeniBook2004}
\bibinfo{author}{\bibfnamefont{S.}~\bibnamefont{Atzeni}} \bibnamefont{and}
  \bibinfo{author}{\bibfnamefont{J.}~\bibnamefont{Meyer-ter Vehn}},
  \emph{\bibinfo{title}{The Physics of Inertial Fusion: BeamPlasma Interaction,
  Hydrodynamics, Hot Dense Matter}} (\bibinfo{publisher}{Oxford University
  Press}, \bibinfo{address}{Oxford}, \bibinfo{year}{2004}).

\bibitem[{\citenamefont{Labaune et~al.}(2013)\citenamefont{Labaune, Baccou,
  Depierreux, Goyon, Loisel, Yahia, and Rafelski}}]{Labaune2013}
\bibinfo{author}{\bibfnamefont{C.}~\bibnamefont{Labaune}},
  \bibinfo{author}{\bibfnamefont{C.}~\bibnamefont{Baccou}},
  \bibinfo{author}{\bibfnamefont{S.}~\bibnamefont{Depierreux}},
  \bibinfo{author}{\bibfnamefont{C.}~\bibnamefont{Goyon}},
  \bibinfo{author}{\bibfnamefont{G.}~\bibnamefont{Loisel}},
  \bibinfo{author}{\bibfnamefont{V.}~\bibnamefont{Yahia}}, \bibnamefont{and}
  \bibinfo{author}{\bibfnamefont{J.}~\bibnamefont{Rafelski}},
  \bibinfo{journal}{Nature Commun.} \textbf{\bibinfo{volume}{4}},
  \bibinfo{pages}{2506} (\bibinfo{year}{2013}).

\bibitem[{\citenamefont{Hurricane et~al.}(2014)\citenamefont{Hurricane,
  Callahan, Casey, Celliers, Cerjan, Dewald, Dittrich, D\"oppner, Hinkel,
  Berzak~Hopkins et~al.}}]{HurricaneNature2014}
\bibinfo{author}{\bibfnamefont{O.~A.} \bibnamefont{Hurricane}},
  \bibinfo{author}{\bibfnamefont{D.~A.} \bibnamefont{Callahan}},
  \bibinfo{author}{\bibfnamefont{D.~T.} \bibnamefont{Casey}},
  \bibinfo{author}{\bibfnamefont{P.~M.} \bibnamefont{Celliers}},
  \bibinfo{author}{\bibfnamefont{C.}~\bibnamefont{Cerjan}},
  \bibinfo{author}{\bibfnamefont{E.~L.} \bibnamefont{Dewald}},
  \bibinfo{author}{\bibfnamefont{T.~R.} \bibnamefont{Dittrich}},
  \bibinfo{author}{\bibfnamefont{T.}~\bibnamefont{D\"oppner}},
  \bibinfo{author}{\bibfnamefont{D.~E.} \bibnamefont{Hinkel}},
  \bibinfo{author}{\bibfnamefont{L.~F.} \bibnamefont{Berzak~Hopkins}},
  \bibnamefont{et~al.}, \bibinfo{journal}{Nature}
  \textbf{\bibinfo{volume}{506}}, \bibinfo{pages}{343} (\bibinfo{year}{2014}).

\bibitem[{\citenamefont{Labaune et~al.}(2016)\citenamefont{Labaune, Baccou,
  Yahia, Neuville, and Rafelski}}]{Labaune2015}
\bibinfo{author}{\bibfnamefont{C.}~\bibnamefont{Labaune}},
  \bibinfo{author}{\bibfnamefont{C.}~\bibnamefont{Baccou}},
  \bibinfo{author}{\bibfnamefont{V.}~\bibnamefont{Yahia}},
  \bibinfo{author}{\bibfnamefont{C.}~\bibnamefont{Neuville}}, \bibnamefont{and}
  \bibinfo{author}{\bibfnamefont{J.}~\bibnamefont{Rafelski}},
  \bibinfo{journal}{Sci. Rep.} \textbf{\bibinfo{volume}{6}},
  \bibinfo{pages}{21202} (\bibinfo{year}{2016}).

\bibitem[{\citenamefont{Olson et~al.}(2016)\citenamefont{Olson, Leeper, Kline,
  Zylstra, Yi, Biener, Braun, Kozioziemski, Sater, Bradley
  et~al.}}]{OlsonPRL2016}
\bibinfo{author}{\bibfnamefont{R.~E.} \bibnamefont{Olson}},
  \bibinfo{author}{\bibfnamefont{R.~J.} \bibnamefont{Leeper}},
  \bibinfo{author}{\bibfnamefont{J.~L.} \bibnamefont{Kline}},
  \bibinfo{author}{\bibfnamefont{A.~B.} \bibnamefont{Zylstra}},
  \bibinfo{author}{\bibfnamefont{S.~A.} \bibnamefont{Yi}},
  \bibinfo{author}{\bibfnamefont{J.}~\bibnamefont{Biener}},
  \bibinfo{author}{\bibfnamefont{T.}~\bibnamefont{Braun}},
  \bibinfo{author}{\bibfnamefont{B.~J.} \bibnamefont{Kozioziemski}},
  \bibinfo{author}{\bibfnamefont{J.~D.} \bibnamefont{Sater}},
  \bibinfo{author}{\bibfnamefont{P.~A.} \bibnamefont{Bradley}},
  \bibnamefont{et~al.}, \bibinfo{journal}{Phys. Rev. Lett.}
  \textbf{\bibinfo{volume}{117}}, \bibinfo{pages}{245001}
  (\bibinfo{year}{2016}).

\bibitem[{\citenamefont{Cerjan et~al.}(2018)\citenamefont{Cerjan, Bernstein,
  Berzak~Hopkins, Bionta, Bleuel, Caggiano, Cassata, Brune, Fittinghoff, Frenje
  et~al.}}]{CerjanJPG2018}
\bibinfo{author}{\bibfnamefont{C.~J.} \bibnamefont{Cerjan}},
  \bibinfo{author}{\bibfnamefont{L.}~\bibnamefont{Bernstein}},
  \bibinfo{author}{\bibfnamefont{L.}~\bibnamefont{Berzak~Hopkins}},
  \bibinfo{author}{\bibfnamefont{R.~M.} \bibnamefont{Bionta}},
  \bibinfo{author}{\bibfnamefont{D.~L.} \bibnamefont{Bleuel}},
  \bibinfo{author}{\bibfnamefont{J.~A.} \bibnamefont{Caggiano}},
  \bibinfo{author}{\bibfnamefont{W.~S.} \bibnamefont{Cassata}},
  \bibinfo{author}{\bibfnamefont{C.~R.} \bibnamefont{Brune}},
  \bibinfo{author}{\bibfnamefont{D.}~\bibnamefont{Fittinghoff}},
  \bibinfo{author}{\bibfnamefont{J.}~\bibnamefont{Frenje}},
  \bibnamefont{et~al.}, \bibinfo{journal}{Journal of Physics G: Nuclear and
  Particle Physics} \textbf{\bibinfo{volume}{45}}, \bibinfo{pages}{033003}
  (\bibinfo{year}{2018}).

\bibitem[{\citenamefont{Salpeter}(1954)}]{Salpeter1954AJP}
\bibinfo{author}{\bibfnamefont{E.~E.} \bibnamefont{Salpeter}},
  \bibinfo{journal}{Australian Journal of Physics}
  \textbf{\bibinfo{volume}{7}}, \bibinfo{pages}{373} (\bibinfo{year}{1954}).

\bibitem[{\citenamefont{Mitler}(1977)}]{MitlerAPJ1977}
\bibinfo{author}{\bibfnamefont{H.~E.} \bibnamefont{Mitler}},
  \bibinfo{journal}{The Astrophysical Journal} \textbf{\bibinfo{volume}{212}},
  \bibinfo{pages}{513} (\bibinfo{year}{1977}).

\bibitem[{\citenamefont{Dzitko et~al.}(1995)\citenamefont{Dzitko,
  Turck-Chi\`eze, Delbourgo-Salvador, and Lagrange}}]{Dzitko1995APJ}
\bibinfo{author}{\bibfnamefont{H.}~\bibnamefont{Dzitko}},
  \bibinfo{author}{\bibfnamefont{S.}~\bibnamefont{Turck-Chi\`eze}},
  \bibinfo{author}{\bibfnamefont{P.}~\bibnamefont{Delbourgo-Salvador}},
  \bibnamefont{and} \bibinfo{author}{\bibfnamefont{C.}~\bibnamefont{Lagrange}},
  \bibinfo{journal}{The Astrophysical Journal} \textbf{\bibinfo{volume}{447}},
  \bibinfo{pages}{428} (\bibinfo{year}{1995}).

\bibitem[{\citenamefont{Salpeter and Van~Horn}(1969)}]{Salpeter1969APJ}
\bibinfo{author}{\bibfnamefont{E.~E.} \bibnamefont{Salpeter}} \bibnamefont{and}
  \bibinfo{author}{\bibfnamefont{H.~M.} \bibnamefont{Van~Horn}},
  \bibinfo{journal}{The Astrophysical Journal} \textbf{\bibinfo{volume}{155}},
  \bibinfo{pages}{183} (\bibinfo{year}{1969}).

\bibitem[{\citenamefont{Dewitt et~al.}(1973)\citenamefont{Dewitt, Graboske, and
  Cooper}}]{DewittAPJ1973}
\bibinfo{author}{\bibfnamefont{H.~E.} \bibnamefont{Dewitt}},
  \bibinfo{author}{\bibfnamefont{H.~C.} \bibnamefont{Graboske}},
  \bibnamefont{and} \bibinfo{author}{\bibfnamefont{M.~S.}
  \bibnamefont{Cooper}}, \bibinfo{journal}{The Astrophysical Journal}
  \textbf{\bibinfo{volume}{181}}, \bibinfo{pages}{439} (\bibinfo{year}{1973}).

\bibitem[{\citenamefont{Graboske et~al.}(1973)\citenamefont{Graboske, Dewitt,
  Grossman, and Cooper}}]{Graboske1973APJ}
\bibinfo{author}{\bibfnamefont{H.~C.} \bibnamefont{Graboske}},
  \bibinfo{author}{\bibfnamefont{H.~E.} \bibnamefont{Dewitt}},
  \bibinfo{author}{\bibfnamefont{A.~S.} \bibnamefont{Grossman}},
  \bibnamefont{and} \bibinfo{author}{\bibfnamefont{M.~S.}
  \bibnamefont{Cooper}}, \bibinfo{journal}{The Astrophysical Journal}
  \textbf{\bibinfo{volume}{181}}, \bibinfo{pages}{457} (\bibinfo{year}{1973}).

\bibitem[{\citenamefont{Gruzinov and Bahcall}(1998)}]{Gruzinov1998APJ}
\bibinfo{author}{\bibfnamefont{A.~V.} \bibnamefont{Gruzinov}} \bibnamefont{and}
  \bibinfo{author}{\bibfnamefont{J.}~\bibnamefont{Bahcall}},
  \bibinfo{journal}{The Astrophysical Journal} \textbf{\bibinfo{volume}{504}},
  \bibinfo{pages}{996} (\bibinfo{year}{1998}).

\bibitem[{\citenamefont{Wilets et~al.}(2000)\citenamefont{Wilets, Giraud,
  Watrous, and Rehr}}]{WiletsAPJ2000}
\bibinfo{author}{\bibfnamefont{L.}~\bibnamefont{Wilets}},
  \bibinfo{author}{\bibfnamefont{B.~G.} \bibnamefont{Giraud}},
  \bibinfo{author}{\bibfnamefont{M.~J.} \bibnamefont{Watrous}},
  \bibnamefont{and} \bibinfo{author}{\bibfnamefont{J.~J.} \bibnamefont{Rehr}},
  \bibinfo{journal}{The Astrophysical Journal} \textbf{\bibinfo{volume}{530}},
  \bibinfo{pages}{504} (\bibinfo{year}{2000}).

\bibitem[{\citenamefont{Bahcall et~al.}(2002)\citenamefont{Bahcall, Brown,
  Gruzinov, and Sawyer}}]{Bahcall2002AA}
\bibinfo{author}{\bibfnamefont{J.~N.} \bibnamefont{Bahcall}},
  \bibinfo{author}{\bibfnamefont{L.~S.} \bibnamefont{Brown}},
  \bibinfo{author}{\bibfnamefont{A.}~\bibnamefont{Gruzinov}}, \bibnamefont{and}
  \bibinfo{author}{\bibfnamefont{R.~F.} \bibnamefont{Sawyer}},
  \bibinfo{journal}{Astronomy \& Astrophysics} \textbf{\bibinfo{volume}{383}},
  \bibinfo{pages}{291} (\bibinfo{year}{2002}).

\bibitem[{\citenamefont{Chitanvis}(2007)}]{Chitanvis2007APJ}
\bibinfo{author}{\bibfnamefont{S.~M.} \bibnamefont{Chitanvis}},
  \bibinfo{journal}{The Astrophysical Journal} \textbf{\bibinfo{volume}{654}},
  \bibinfo{pages}{693} (\bibinfo{year}{2007}).

\bibitem[{\citenamefont{Ciricosta et~al.}(2012)\citenamefont{Ciricosta, Vinko,
  Chung, Cho, Brown, Burian, Chalupsk\'y, Engelhorn, Falcone, Graves
  et~al.}}]{CiricostaPRL2012}
\bibinfo{author}{\bibfnamefont{O.}~\bibnamefont{Ciricosta}},
  \bibinfo{author}{\bibfnamefont{S.~M.} \bibnamefont{Vinko}},
  \bibinfo{author}{\bibfnamefont{H.-K.} \bibnamefont{Chung}},
  \bibinfo{author}{\bibfnamefont{B.-I.} \bibnamefont{Cho}},
  \bibinfo{author}{\bibfnamefont{C.~R.~D.} \bibnamefont{Brown}},
  \bibinfo{author}{\bibfnamefont{T.}~\bibnamefont{Burian}},
  \bibinfo{author}{\bibfnamefont{J.}~\bibnamefont{Chalupsk\'y}},
  \bibinfo{author}{\bibfnamefont{K.}~\bibnamefont{Engelhorn}},
  \bibinfo{author}{\bibfnamefont{R.~W.} \bibnamefont{Falcone}},
  \bibinfo{author}{\bibfnamefont{C.}~\bibnamefont{Graves}},
  \bibnamefont{et~al.}, \bibinfo{journal}{Phys. Rev. Lett.}
  \textbf{\bibinfo{volume}{109}}, \bibinfo{pages}{065002}
  (\bibinfo{year}{2012}).

\bibitem[{\citenamefont{Stewart and Pyatt}(1966)}]{StewartAPJ1966}
\bibinfo{author}{\bibfnamefont{J.~C.} \bibnamefont{Stewart}} \bibnamefont{and}
  \bibinfo{author}{\bibfnamefont{K.~D.} \bibnamefont{Pyatt},
  \bibfnamefont{Jr.}}, \bibinfo{journal}{The Astrophysical Journal}
  \textbf{\bibinfo{volume}{144}}, \bibinfo{pages}{1203} (\bibinfo{year}{1966}).

\bibitem[{\citenamefont{Debye and H\"{u}ckel}(1923)}]{DebyePZ1923}
\bibinfo{author}{\bibfnamefont{P.}~\bibnamefont{Debye}} \bibnamefont{and}
  \bibinfo{author}{\bibfnamefont{E.}~\bibnamefont{H\"{u}ckel}},
  \bibinfo{journal}{Physikalische Zeitschrift} \textbf{\bibinfo{volume}{24}},
  \bibinfo{pages}{185} (\bibinfo{year}{1923}).

\bibitem[{\citenamefont{Griem}(1964)}]{GriemBook1964}
\bibinfo{author}{\bibfnamefont{H.~R.} \bibnamefont{Griem}},
  \emph{\bibinfo{title}{Plasma Spectroscopy}}
  (\bibinfo{publisher}{McGraw-Hill}, \bibinfo{address}{New York},
  \bibinfo{year}{1964}).

\bibitem[{\citenamefont{Hoarty et~al.}(2013)\citenamefont{Hoarty, Allan, James,
  Brown, Hobbs, Hill, Harris, Morton, Brookes, Shepherd
  et~al.}}]{HoartyPRL2013}
\bibinfo{author}{\bibfnamefont{D.~J.} \bibnamefont{Hoarty}},
  \bibinfo{author}{\bibfnamefont{P.}~\bibnamefont{Allan}},
  \bibinfo{author}{\bibfnamefont{S.~F.} \bibnamefont{James}},
  \bibinfo{author}{\bibfnamefont{C.~R.~D.} \bibnamefont{Brown}},
  \bibinfo{author}{\bibfnamefont{L.~M.~R.} \bibnamefont{Hobbs}},
  \bibinfo{author}{\bibfnamefont{M.~P.} \bibnamefont{Hill}},
  \bibinfo{author}{\bibfnamefont{J.~W.~O.} \bibnamefont{Harris}},
  \bibinfo{author}{\bibfnamefont{J.}~\bibnamefont{Morton}},
  \bibinfo{author}{\bibfnamefont{M.~G.} \bibnamefont{Brookes}},
  \bibinfo{author}{\bibfnamefont{R.}~\bibnamefont{Shepherd}},
  \bibnamefont{et~al.}, \bibinfo{journal}{Phys. Rev. Lett.}
  \textbf{\bibinfo{volume}{110}}, \bibinfo{pages}{265003}
  (\bibinfo{year}{2013}).

\bibitem[{\citenamefont{Fletcher et~al.}(2014)\citenamefont{Fletcher, Kritcher,
  Pak, Ma, D\"oppner, Fortmann, Divol, Jones, Landen, Scott
  et~al.}}]{FletcherPRL2014}
\bibinfo{author}{\bibfnamefont{L.~B.} \bibnamefont{Fletcher}},
  \bibinfo{author}{\bibfnamefont{A.~L.} \bibnamefont{Kritcher}},
  \bibinfo{author}{\bibfnamefont{A.}~\bibnamefont{Pak}},
  \bibinfo{author}{\bibfnamefont{T.}~\bibnamefont{Ma}},
  \bibinfo{author}{\bibfnamefont{T.}~\bibnamefont{D\"oppner}},
  \bibinfo{author}{\bibfnamefont{C.}~\bibnamefont{Fortmann}},
  \bibinfo{author}{\bibfnamefont{L.}~\bibnamefont{Divol}},
  \bibinfo{author}{\bibfnamefont{O.~S.} \bibnamefont{Jones}},
  \bibinfo{author}{\bibfnamefont{O.~L.} \bibnamefont{Landen}},
  \bibinfo{author}{\bibfnamefont{H.~A.} \bibnamefont{Scott}},
  \bibnamefont{et~al.}, \bibinfo{journal}{Phys. Rev. Lett.}
  \textbf{\bibinfo{volume}{112}}, \bibinfo{pages}{145004}
  (\bibinfo{year}{2014}).

\bibitem[{\citenamefont{Ciricosta et~al.}(2016)\citenamefont{Ciricosta, Vinko,
  Barbrel, Rackstraw, Preston, Burian, Chalupsk{\'y}, Cho, Chung, Dakovski
  et~al.}}]{CiricostaNC2016}
\bibinfo{author}{\bibfnamefont{O.}~\bibnamefont{Ciricosta}},
  \bibinfo{author}{\bibfnamefont{S.~M.} \bibnamefont{Vinko}},
  \bibinfo{author}{\bibfnamefont{B.}~\bibnamefont{Barbrel}},
  \bibinfo{author}{\bibfnamefont{D.~S.} \bibnamefont{Rackstraw}},
  \bibinfo{author}{\bibfnamefont{T.~R.} \bibnamefont{Preston}},
  \bibinfo{author}{\bibfnamefont{T.}~\bibnamefont{Burian}},
  \bibinfo{author}{\bibfnamefont{J.}~\bibnamefont{Chalupsk{\'y}}},
  \bibinfo{author}{\bibfnamefont{B.~I.} \bibnamefont{Cho}},
  \bibinfo{author}{\bibfnamefont{H.~K.} \bibnamefont{Chung}},
  \bibinfo{author}{\bibfnamefont{G.~L.} \bibnamefont{Dakovski}},
  \bibnamefont{et~al.}, \bibinfo{journal}{Nature Communications}
  \textbf{\bibinfo{volume}{7}}, \bibinfo{pages}{11713} (\bibinfo{year}{2016}).

\bibitem[{\citenamefont{Kraus et~al.}(2019)\citenamefont{Kraus, Bachmann,
  Barbrel, Falcone, Fletcher, Frydrych, Gamboa, Gauthier, Gericke, Glenzer
  et~al.}}]{KrausPPCF2019}
\bibinfo{author}{\bibfnamefont{D.}~\bibnamefont{Kraus}},
  \bibinfo{author}{\bibfnamefont{B.}~\bibnamefont{Bachmann}},
  \bibinfo{author}{\bibfnamefont{B.}~\bibnamefont{Barbrel}},
  \bibinfo{author}{\bibfnamefont{R.~W.} \bibnamefont{Falcone}},
  \bibinfo{author}{\bibfnamefont{L.~B.} \bibnamefont{Fletcher}},
  \bibinfo{author}{\bibfnamefont{S.}~\bibnamefont{Frydrych}},
  \bibinfo{author}{\bibfnamefont{E.~J.} \bibnamefont{Gamboa}},
  \bibinfo{author}{\bibfnamefont{M.}~\bibnamefont{Gauthier}},
  \bibinfo{author}{\bibfnamefont{D.~O.} \bibnamefont{Gericke}},
  \bibinfo{author}{\bibfnamefont{S.~H.} \bibnamefont{Glenzer}},
  \bibnamefont{et~al.}, \bibinfo{journal}{Plasma Physics and Controlled Fusion}
  \textbf{\bibinfo{volume}{61}}, \bibinfo{pages}{014015}
  (\bibinfo{year}{2019}).

\bibitem[{\citenamefont{Vinko et~al.}(2014)\citenamefont{Vinko, Ciricosta, and
  Wark}}]{VinkoNC2014}
\bibinfo{author}{\bibfnamefont{S.~M.} \bibnamefont{Vinko}},
  \bibinfo{author}{\bibfnamefont{O.}~\bibnamefont{Ciricosta}},
  \bibnamefont{and} \bibinfo{author}{\bibfnamefont{J.~S.} \bibnamefont{Wark}},
  \bibinfo{journal}{Nature Communications} \textbf{\bibinfo{volume}{5}},
  \bibinfo{pages}{3533} (\bibinfo{year}{2014}).

\bibitem[{\citenamefont{Jin et~al.}(2021)\citenamefont{Jin, Abdullah, Jurek,
  Santra, and Son}}]{JinPRE2021}
\bibinfo{author}{\bibfnamefont{R.}~\bibnamefont{Jin}},
  \bibinfo{author}{\bibfnamefont{M.~M.} \bibnamefont{Abdullah}},
  \bibinfo{author}{\bibfnamefont{Z.}~\bibnamefont{Jurek}},
  \bibinfo{author}{\bibfnamefont{R.}~\bibnamefont{Santra}}, \bibnamefont{and}
  \bibinfo{author}{\bibfnamefont{S.-K.} \bibnamefont{Son}},
  \bibinfo{journal}{Phys. Rev. E} \textbf{\bibinfo{volume}{103}},
  \bibinfo{pages}{023203} (\bibinfo{year}{2021}).

\bibitem[{\citenamefont{Zeng et~al.}(2020)\citenamefont{Zeng, Li, Gao, and
  Yuan}}]{ZengAA2020}
\bibinfo{author}{\bibfnamefont{J.}~\bibnamefont{Zeng}},
  \bibinfo{author}{\bibfnamefont{Y.}~\bibnamefont{Li}},
  \bibinfo{author}{\bibfnamefont{C.}~\bibnamefont{Gao}}, \bibnamefont{and}
  \bibinfo{author}{\bibfnamefont{J.}~\bibnamefont{Yuan}},
  \bibinfo{journal}{Astronomy \& Astrophysics} \textbf{\bibinfo{volume}{634}},
  \bibinfo{pages}{A117} (\bibinfo{year}{2020}).

\bibitem[{\citenamefont{Tsytovich and Bornatici}(2000)}]{Tsytovich2000}
\bibinfo{author}{\bibfnamefont{V.~N.} \bibnamefont{Tsytovich}}
  \bibnamefont{and}
  \bibinfo{author}{\bibfnamefont{M.}~\bibnamefont{Bornatici}},
  \bibinfo{journal}{Plasma Phys. Rep.} \textbf{\bibinfo{volume}{26}},
  \bibinfo{pages}{840} (\bibinfo{year}{2000}).

\bibitem[{\citenamefont{Negoita et~al.}(2016)\citenamefont{Negoita, Roth,
  Thirolf, Tudisco, Hannachi, Moustaizis, Pomerantz, Mckenna, Fuchs, Spohr
  et~al.}}]{NegoitaRRP2016}
\bibinfo{author}{\bibfnamefont{F.}~\bibnamefont{Negoita}},
  \bibinfo{author}{\bibfnamefont{M.}~\bibnamefont{Roth}},
  \bibinfo{author}{\bibfnamefont{P.~G.} \bibnamefont{Thirolf}},
  \bibinfo{author}{\bibfnamefont{S.}~\bibnamefont{Tudisco}},
  \bibinfo{author}{\bibfnamefont{F.}~\bibnamefont{Hannachi}},
  \bibinfo{author}{\bibfnamefont{S.}~\bibnamefont{Moustaizis}},
  \bibinfo{author}{\bibfnamefont{I.}~\bibnamefont{Pomerantz}},
  \bibinfo{author}{\bibfnamefont{P.}~\bibnamefont{Mckenna}},
  \bibinfo{author}{\bibfnamefont{J.}~\bibnamefont{Fuchs}},
  \bibinfo{author}{\bibfnamefont{K.}~\bibnamefont{Spohr}},
  \bibnamefont{et~al.}, \bibinfo{journal}{Romanian Reports in Physics}
  \textbf{\bibinfo{volume}{68, Supplement}}, \bibinfo{pages}{S37}
  (\bibinfo{year}{2016}).

\bibitem[{\citenamefont{Keller}(1953)}]{Keller1953APJ}
\bibinfo{author}{\bibfnamefont{G.}~\bibnamefont{Keller}}, \bibinfo{journal}{The
  Astrophysical Journal} \textbf{\bibinfo{volume}{118}}, \bibinfo{pages}{142}
  (\bibinfo{year}{1953}).

\bibitem[{\citenamefont{Brown and Sawyer}(1997)}]{BrownRMP1997}
\bibinfo{author}{\bibfnamefont{L.~S.} \bibnamefont{Brown}} \bibnamefont{and}
  \bibinfo{author}{\bibfnamefont{R.~F.} \bibnamefont{Sawyer}},
  \bibinfo{journal}{Rev. Mod. Phys.} \textbf{\bibinfo{volume}{69}},
  \bibinfo{pages}{411} (\bibinfo{year}{1997}).

\bibitem[{\citenamefont{Gruzinov}(1998)}]{GruzinovAVApJ1998b}
\bibinfo{author}{\bibfnamefont{A.~V.} \bibnamefont{Gruzinov}},
  \bibinfo{journal}{The Astrophysical Journal} \textbf{\bibinfo{volume}{496}},
  \bibinfo{pages}{503} (\bibinfo{year}{1998}).

\bibitem[{\citenamefont{Kravchuk and Yakovlev}(2014)}]{KravchukPRC2014}
\bibinfo{author}{\bibfnamefont{P.~A.} \bibnamefont{Kravchuk}} \bibnamefont{and}
  \bibinfo{author}{\bibfnamefont{D.~G.} \bibnamefont{Yakovlev}},
  \bibinfo{journal}{Phys. Rev. C} \textbf{\bibinfo{volume}{89}},
  \bibinfo{pages}{015802} (\bibinfo{year}{2014}).

\bibitem[{\citenamefont{Shaviv and Shaviv}(1996)}]{Shaviv1996APJ}
\bibinfo{author}{\bibfnamefont{N.~J.} \bibnamefont{Shaviv}} \bibnamefont{and}
  \bibinfo{author}{\bibfnamefont{G.}~\bibnamefont{Shaviv}},
  \bibinfo{journal}{The Astrophysical Journal} \textbf{\bibinfo{volume}{468}},
  \bibinfo{pages}{433} (\bibinfo{year}{1996}).

\bibitem[{\citenamefont{Shaviv and Shaviv}(2000)}]{Shaviv2000APJ}
\bibinfo{author}{\bibfnamefont{G.}~\bibnamefont{Shaviv}} \bibnamefont{and}
  \bibinfo{author}{\bibfnamefont{N.~J.} \bibnamefont{Shaviv}},
  \bibinfo{journal}{The Astrophysical Journal} \textbf{\bibinfo{volume}{529}},
  \bibinfo{pages}{1054} (\bibinfo{year}{2000}).

\bibitem[{\citenamefont{Mao et~al.}(2009)\citenamefont{Mao, Mussack, and
  D\"appen}}]{Mao2009APJ}
\bibinfo{author}{\bibfnamefont{D.}~\bibnamefont{Mao}},
  \bibinfo{author}{\bibfnamefont{K.}~\bibnamefont{Mussack}}, \bibnamefont{and}
  \bibinfo{author}{\bibfnamefont{W.}~\bibnamefont{D\"appen}},
  \bibinfo{journal}{The Astrophysical Journal} \textbf{\bibinfo{volume}{701}},
  \bibinfo{pages}{1204} (\bibinfo{year}{2009}).

\bibitem[{\citenamefont{Kushnir et~al.}(2019)\citenamefont{Kushnir, Waxman, and
  Chugunov}}]{KushnirMNRAS2019}
\bibinfo{author}{\bibfnamefont{D.}~\bibnamefont{Kushnir}},
  \bibinfo{author}{\bibfnamefont{E.}~\bibnamefont{Waxman}}, \bibnamefont{and}
  \bibinfo{author}{\bibfnamefont{A.~I.} \bibnamefont{Chugunov}},
  \bibinfo{journal}{Monthly Notices of the Royal Astronomical Society}
  \textbf{\bibinfo{volume}{486}}, \bibinfo{pages}{449} (\bibinfo{year}{2019}).

\bibitem[{\citenamefont{Cl{\'e}rouin et~al.}(2019)\citenamefont{Cl{\'e}rouin,
  Arnault, Desbiens, White, Collins, Kress, and Ticknor}}]{ClerouinPoP2019}
\bibinfo{author}{\bibfnamefont{J.}~\bibnamefont{Cl{\'e}rouin}},
  \bibinfo{author}{\bibfnamefont{P.}~\bibnamefont{Arnault}},
  \bibinfo{author}{\bibfnamefont{N.}~\bibnamefont{Desbiens}},
  \bibinfo{author}{\bibfnamefont{A.~J.} \bibnamefont{White}},
  \bibinfo{author}{\bibfnamefont{L.~A.} \bibnamefont{Collins}},
  \bibinfo{author}{\bibfnamefont{J.~D.} \bibnamefont{Kress}}, \bibnamefont{and}
  \bibinfo{author}{\bibfnamefont{C.}~\bibnamefont{Ticknor}},
  \bibinfo{journal}{Physics of Plasmas} \textbf{\bibinfo{volume}{26}},
  \bibinfo{pages}{012702} (\bibinfo{year}{2019}).

\bibitem[{\citenamefont{{European XFEL}}(2022)}]{europeanXFEL-web}
\bibinfo{author}{\bibnamefont{{European XFEL}}},
  \bibinfo{howpublished}{Official Website} (\bibinfo{year}{2022}),
  \bibinfo{note}{{ {https://www.xfel.eu/}}}.

\bibitem[{\citenamefont{{Linac Coherent Light Source}}(2022)}]{LCLS-web}
\bibinfo{author}{\bibnamefont{{Linac Coherent Light Source}}},
  \bibinfo{howpublished}{Official Website} (\bibinfo{year}{2022}),
  \bibinfo{note}{{ {https://lcls.slac.stanford.edu/}}}.

\bibitem[{\citenamefont{{SwissFEL}}(2022)}]{SwissFEL-web}
\bibinfo{author}{\bibnamefont{{SwissFEL}}}, \bibinfo{howpublished}{Official
  Website} (\bibinfo{year}{2022}), \bibinfo{note}{{
  {https://www.psi.ch/en/swissfel}}}.

\bibitem[{\citenamefont{{SACLA}}(2022)}]{SACLA-web}
\bibinfo{author}{\bibnamefont{{SACLA}}}, \bibinfo{howpublished}{Official
  Website} (\bibinfo{year}{2022}), \bibinfo{note}{{
  {http://xfel.riken.jp/eng/}}}.

\bibitem[{\citenamefont{Danson et~al.}(2015)\citenamefont{Danson, Hillier,
  Hopps, and Neely}}]{DansonHPLSE2014}
\bibinfo{author}{\bibfnamefont{C.}~\bibnamefont{Danson}},
  \bibinfo{author}{\bibfnamefont{D.}~\bibnamefont{Hillier}},
  \bibinfo{author}{\bibfnamefont{N.}~\bibnamefont{Hopps}}, \bibnamefont{and}
  \bibinfo{author}{\bibfnamefont{D.}~\bibnamefont{Neely}},
  \bibinfo{journal}{High Power Laser Science and Engineering}
  \textbf{\bibinfo{volume}{3}}, \bibinfo{pages}{e3} (\bibinfo{year}{2015}).

\bibitem[{\citenamefont{Danson et~al.}(2019)\citenamefont{Danson, Haefner,
  Bromage, Butcher, Chanteloup, Chowdhury, Galvanauskas, Gizzi, Hein, Hillier
  et~al.}}]{DansonHPLSE2019}
\bibinfo{author}{\bibfnamefont{C.~N.} \bibnamefont{Danson}},
  \bibinfo{author}{\bibfnamefont{C.}~\bibnamefont{Haefner}},
  \bibinfo{author}{\bibfnamefont{J.}~\bibnamefont{Bromage}},
  \bibinfo{author}{\bibfnamefont{T.}~\bibnamefont{Butcher}},
  \bibinfo{author}{\bibfnamefont{J.-C.~F.} \bibnamefont{Chanteloup}},
  \bibinfo{author}{\bibfnamefont{E.~A.} \bibnamefont{Chowdhury}},
  \bibinfo{author}{\bibfnamefont{A.}~\bibnamefont{Galvanauskas}},
  \bibinfo{author}{\bibfnamefont{L.~A.} \bibnamefont{Gizzi}},
  \bibinfo{author}{\bibfnamefont{J.}~\bibnamefont{Hein}},
  \bibinfo{author}{\bibfnamefont{D.~I.} \bibnamefont{Hillier}},
  \bibnamefont{et~al.}, \bibinfo{journal}{High Power Laser Science and
  Engineering} \textbf{\bibinfo{volume}{7}}, \bibinfo{pages}{e54}
  (\bibinfo{year}{2019}).

\bibitem[{\citenamefont{{Extreme Light Infrastructure (ELI)}}(2022)}]{ELI-web}
\bibinfo{author}{\bibnamefont{{Extreme Light Infrastructure (ELI)}}},
  \bibinfo{howpublished}{Official Website} (\bibinfo{year}{2022}),
  \bibinfo{note}{{ {https://eli-laser.eu/}}}.

\bibitem[{\citenamefont{{LULI}}(2022)}]{LULI-web}
\bibinfo{author}{\bibnamefont{{LULI}}}, \bibinfo{howpublished}{Official
  Website} (\bibinfo{year}{2022}), \bibinfo{note}{{
  {https://portail.polytechnique.edu/luli/en}}}.

\bibitem[{\citenamefont{{Vulcan laser}}(2022)}]{Vulcan-web}
\bibinfo{author}{\bibnamefont{{Vulcan laser}}}, \bibinfo{howpublished}{Official
  Website} (\bibinfo{year}{2022}), \bibinfo{note}{{
  {https://www.clf.stfc.ac.uk/Pages/Vulcan-laser.aspx}}}.

\bibitem[{\citenamefont{{National Ignition Facility (NIF)}}(2022)}]{NIF-web}
\bibinfo{author}{\bibnamefont{{National Ignition Facility (NIF)}}},
  \bibinfo{howpublished}{Official Website} (\bibinfo{year}{2022}),
  \bibinfo{note}{{ {https://lasers.llnl.gov/}}}.

\bibitem[{\citenamefont{Casey et~al.}(2017)\citenamefont{Casey, Sayre, Brune,
  Smalyuk, Weber, Tipton, Pino, Grim, Remington, Dearborn
  et~al.}}]{CaseyNP2017}
\bibinfo{author}{\bibfnamefont{D.~T.} \bibnamefont{Casey}},
  \bibinfo{author}{\bibfnamefont{D.~B.} \bibnamefont{Sayre}},
  \bibinfo{author}{\bibfnamefont{C.~R.} \bibnamefont{Brune}},
  \bibinfo{author}{\bibfnamefont{V.~A.} \bibnamefont{Smalyuk}},
  \bibinfo{author}{\bibfnamefont{C.~R.} \bibnamefont{Weber}},
  \bibinfo{author}{\bibfnamefont{R.~E.} \bibnamefont{Tipton}},
  \bibinfo{author}{\bibfnamefont{J.~E.} \bibnamefont{Pino}},
  \bibinfo{author}{\bibfnamefont{G.~P.} \bibnamefont{Grim}},
  \bibinfo{author}{\bibfnamefont{B.~A.} \bibnamefont{Remington}},
  \bibinfo{author}{\bibfnamefont{D.}~\bibnamefont{Dearborn}},
  \bibnamefont{et~al.}, \bibinfo{journal}{Nature Physics}
  \textbf{\bibinfo{volume}{13}}, \bibinfo{pages}{1227} (\bibinfo{year}{2017}).

\bibitem[{\citenamefont{Gunst et~al.}(2014)\citenamefont{Gunst, Litvinov,
  Keitel, and P\'alffy}}]{GunstPRL2014}
\bibinfo{author}{\bibfnamefont{J.}~\bibnamefont{Gunst}},
  \bibinfo{author}{\bibfnamefont{Y.~A.} \bibnamefont{Litvinov}},
  \bibinfo{author}{\bibfnamefont{C.~H.} \bibnamefont{Keitel}},
  \bibnamefont{and} \bibinfo{author}{\bibfnamefont{A.}~\bibnamefont{P\'alffy}},
  \bibinfo{journal}{Phys. Rev. Lett.} \textbf{\bibinfo{volume}{112}},
  \bibinfo{pages}{082501} (\bibinfo{year}{2014}).

\bibitem[{\citenamefont{Gunst et~al.}(2015)\citenamefont{Gunst, Wu, Kumar,
  Keitel, and P\'alffy}}]{GunstPOP2015}
\bibinfo{author}{\bibfnamefont{J.}~\bibnamefont{Gunst}},
  \bibinfo{author}{\bibfnamefont{Y.}~\bibnamefont{Wu}},
  \bibinfo{author}{\bibfnamefont{N.}~\bibnamefont{Kumar}},
  \bibinfo{author}{\bibfnamefont{C.~H.} \bibnamefont{Keitel}},
  \bibnamefont{and} \bibinfo{author}{\bibfnamefont{A.}~\bibnamefont{P\'alffy}},
  \bibinfo{journal}{Physics of Plasmas} \textbf{\bibinfo{volume}{22}},
  \bibinfo{pages}{112706} (\bibinfo{year}{2015}).

\bibitem[{\citenamefont{Wu and P\'alffy}(2017)}]{Wu2017APJ}
\bibinfo{author}{\bibfnamefont{Y.}~\bibnamefont{Wu}} \bibnamefont{and}
  \bibinfo{author}{\bibfnamefont{A.}~\bibnamefont{P\'alffy}},
  \bibinfo{journal}{The Astrophysical Journal} \textbf{\bibinfo{volume}{838}},
  \bibinfo{pages}{55} (\bibinfo{year}{2017}).

\bibitem[{\citenamefont{Wu}(2020)}]{WuPOP2020}
\bibinfo{author}{\bibfnamefont{Y.}~\bibnamefont{Wu}}, \bibinfo{journal}{Physics
  of Plasmas} \textbf{\bibinfo{volume}{27}}, \bibinfo{pages}{022708}
  (\bibinfo{year}{2020}).

\bibitem[{\citenamefont{Paxton et~al.}(2011)\citenamefont{Paxton, Bildsten,
  Dotter, Herwig, Lesaffre, and Timmes}}]{PaxtonAPJS2011}
\bibinfo{author}{\bibfnamefont{B.}~\bibnamefont{Paxton}},
  \bibinfo{author}{\bibfnamefont{L.}~\bibnamefont{Bildsten}},
  \bibinfo{author}{\bibfnamefont{A.}~\bibnamefont{Dotter}},
  \bibinfo{author}{\bibfnamefont{F.}~\bibnamefont{Herwig}},
  \bibinfo{author}{\bibfnamefont{P.}~\bibnamefont{Lesaffre}}, \bibnamefont{and}
  \bibinfo{author}{\bibfnamefont{F.}~\bibnamefont{Timmes}},
  \bibinfo{journal}{The Astrophysical Journal Supplement Series}
  \textbf{\bibinfo{volume}{192}}, \bibinfo{pages}{3} (\bibinfo{year}{2011}).

\bibitem[{\citenamefont{Hidalgo et~al.}(2018)\citenamefont{Hidalgo,
  Pietrinferni, Cassisi, Salaris, Mucciarelli, Savino, Aparicio, Silva~Aguirre,
  and Verma}}]{HidalgoAPJ2018}
\bibinfo{author}{\bibfnamefont{S.~L.} \bibnamefont{Hidalgo}},
  \bibinfo{author}{\bibfnamefont{A.}~\bibnamefont{Pietrinferni}},
  \bibinfo{author}{\bibfnamefont{S.}~\bibnamefont{Cassisi}},
  \bibinfo{author}{\bibfnamefont{M.}~\bibnamefont{Salaris}},
  \bibinfo{author}{\bibfnamefont{A.}~\bibnamefont{Mucciarelli}},
  \bibinfo{author}{\bibfnamefont{A.}~\bibnamefont{Savino}},
  \bibinfo{author}{\bibfnamefont{A.}~\bibnamefont{Aparicio}},
  \bibinfo{author}{\bibfnamefont{V.}~\bibnamefont{Silva~Aguirre}},
  \bibnamefont{and} \bibinfo{author}{\bibfnamefont{K.}~\bibnamefont{Verma}},
  \bibinfo{journal}{The Astrophysical Journal} \textbf{\bibinfo{volume}{856}},
  \bibinfo{pages}{125} (\bibinfo{year}{2018}).

\bibitem[{\citenamefont{Forbes and Loeb}(2019)}]{ForbesAPJ2019}
\bibinfo{author}{\bibfnamefont{J.~C.} \bibnamefont{Forbes}} \bibnamefont{and}
  \bibinfo{author}{\bibfnamefont{A.}~\bibnamefont{Loeb}}, \bibinfo{journal}{The
  Astrophysical Journal} \textbf{\bibinfo{volume}{871}}, \bibinfo{pages}{227}
  (\bibinfo{year}{2019}).

\bibitem[{\citenamefont{Moussa}(2017)}]{MoussaEPL2017}
\bibinfo{author}{\bibfnamefont{M.}~\bibnamefont{Moussa}},
  \bibinfo{journal}{Europhysics Letters} \textbf{\bibinfo{volume}{117}},
  \bibinfo{pages}{49002} (\bibinfo{year}{2017}).

\bibitem[{\citenamefont{Aguilera-G{\'o}mez
  et~al.}(2016)\citenamefont{Aguilera-G{\'o}mez, Chanam{\'e}, Pinsonneault, and
  Carlberg}}]{AguileraAPJ2016}
\bibinfo{author}{\bibfnamefont{C.}~\bibnamefont{Aguilera-G{\'o}mez}},
  \bibinfo{author}{\bibfnamefont{J.}~\bibnamefont{Chanam{\'e}}},
  \bibinfo{author}{\bibfnamefont{M.~H.} \bibnamefont{Pinsonneault}},
  \bibnamefont{and} \bibinfo{author}{\bibfnamefont{J.~K.}
  \bibnamefont{Carlberg}}, \bibinfo{journal}{The Astrophysical Journal}
  \textbf{\bibinfo{volume}{829}}, \bibinfo{pages}{127} (\bibinfo{year}{2016}).

\bibitem[{\citenamefont{Somers and Pinsonneault}(2016)}]{SomersAPJ2016}
\bibinfo{author}{\bibfnamefont{G.}~\bibnamefont{Somers}} \bibnamefont{and}
  \bibinfo{author}{\bibfnamefont{M.~H.} \bibnamefont{Pinsonneault}},
  \bibinfo{journal}{The Astrophysical Journal} \textbf{\bibinfo{volume}{829}},
  \bibinfo{pages}{32} (\bibinfo{year}{2016}).

\bibitem[{\citenamefont{Tognelli et~al.}(2015)\citenamefont{Tognelli,
  Prada~Moroni, and Degl'Innocenti}}]{TognelliMNRAS2015}
\bibinfo{author}{\bibfnamefont{E.}~\bibnamefont{Tognelli}},
  \bibinfo{author}{\bibfnamefont{P.~G.} \bibnamefont{Prada~Moroni}},
  \bibnamefont{and}
  \bibinfo{author}{\bibfnamefont{S.}~\bibnamefont{Degl'Innocenti}},
  \bibinfo{journal}{Monthly Notices of the Royal Astronomical Society}
  \textbf{\bibinfo{volume}{449}}, \bibinfo{pages}{3741} (\bibinfo{year}{2015}).

\bibitem[{\citenamefont{Young et~al.}(2003)\citenamefont{Young, Knierman,
  Rigby, and Arnett}}]{YoungAPJ2003}
\bibinfo{author}{\bibfnamefont{P.~A.} \bibnamefont{Young}},
  \bibinfo{author}{\bibfnamefont{K.~A.} \bibnamefont{Knierman}},
  \bibinfo{author}{\bibfnamefont{J.~R.} \bibnamefont{Rigby}}, \bibnamefont{and}
  \bibinfo{author}{\bibfnamefont{D.}~\bibnamefont{Arnett}},
  \bibinfo{journal}{The Astrophysical Journal} \textbf{\bibinfo{volume}{595}},
  \bibinfo{pages}{1114} (\bibinfo{year}{2003}).

\bibitem[{\citenamefont{Chabrier and Baraffe}(2000)}]{ChabrierARAA2000}
\bibinfo{author}{\bibfnamefont{G.}~\bibnamefont{Chabrier}} \bibnamefont{and}
  \bibinfo{author}{\bibfnamefont{I.}~\bibnamefont{Baraffe}},
  \bibinfo{journal}{Annual Review of Astronomy and Astrophysics}
  \textbf{\bibinfo{volume}{38}}, \bibinfo{pages}{337} (\bibinfo{year}{2000}).

\bibitem[{\citenamefont{Chabrier and Baraffe}(1997)}]{ChabrierAA1997}
\bibinfo{author}{\bibfnamefont{G.}~\bibnamefont{Chabrier}} \bibnamefont{and}
  \bibinfo{author}{\bibfnamefont{I.}~\bibnamefont{Baraffe}},
  \bibinfo{journal}{Astronomy and Astrophysics} \textbf{\bibinfo{volume}{327}},
  \bibinfo{pages}{1039} (\bibinfo{year}{1997}).

\bibitem[{\citenamefont{Lagarde et~al.}(2012)\citenamefont{Lagarde, Decressin,
  Charbonnel, Eggenberger, Ekstr{\"o}m, and Palacios}}]{LagardeAA2012}
\bibinfo{author}{\bibfnamefont{N.}~\bibnamefont{Lagarde}},
  \bibinfo{author}{\bibfnamefont{T.}~\bibnamefont{Decressin}},
  \bibinfo{author}{\bibfnamefont{C.}~\bibnamefont{Charbonnel}},
  \bibinfo{author}{\bibfnamefont{P.}~\bibnamefont{Eggenberger}},
  \bibinfo{author}{\bibfnamefont{S.}~\bibnamefont{Ekstr{\"o}m}},
  \bibnamefont{and} \bibinfo{author}{\bibfnamefont{A.}~\bibnamefont{Palacios}},
  \bibinfo{journal}{Astronomy \& Astrophysics} \textbf{\bibinfo{volume}{543}},
  \bibinfo{pages}{A108} (\bibinfo{year}{2012}).

\bibitem[{\citenamefont{Amard et~al.}(2019)\citenamefont{Amard, Palacios,
  Charbonnel, Gallet, Georgy, Lagarde, and Siess}}]{AmardAA2019}
\bibinfo{author}{\bibfnamefont{L.}~\bibnamefont{Amard}},
  \bibinfo{author}{\bibfnamefont{A.}~\bibnamefont{Palacios}},
  \bibinfo{author}{\bibfnamefont{C.}~\bibnamefont{Charbonnel}},
  \bibinfo{author}{\bibfnamefont{F.}~\bibnamefont{Gallet}},
  \bibinfo{author}{\bibfnamefont{C.}~\bibnamefont{Georgy}},
  \bibinfo{author}{\bibfnamefont{N.}~\bibnamefont{Lagarde}}, \bibnamefont{and}
  \bibinfo{author}{\bibfnamefont{L.}~\bibnamefont{Siess}},
  \bibinfo{journal}{Astronomy \& Astrophysics} \textbf{\bibinfo{volume}{631}},
  \bibinfo{pages}{A77} (\bibinfo{year}{2019}).

\bibitem[{\citenamefont{de~Loore C. W.~H. and C.}(1992)}]{DeLooreBook1992}
\bibinfo{author}{\bibnamefont{de~Loore C. W.~H.}} \bibnamefont{and}
  \bibinfo{author}{\bibfnamefont{D.}~\bibnamefont{C.}},
  \emph{\bibinfo{title}{Structure and Evolution of Single and Binary Stars}}
  (\bibinfo{publisher}{Springer}, \bibinfo{address}{Dordrecht},
  \bibinfo{year}{1992}).

\bibitem[{\citenamefont{Gallino et~al.}(1998)\citenamefont{Gallino, Arlandini,
  Busso, Lugaro, Travaglio, Straniero, Chieffi, and Limongi}}]{GallinoAPJ1998}
\bibinfo{author}{\bibfnamefont{R.}~\bibnamefont{Gallino}},
  \bibinfo{author}{\bibfnamefont{C.}~\bibnamefont{Arlandini}},
  \bibinfo{author}{\bibfnamefont{M.}~\bibnamefont{Busso}},
  \bibinfo{author}{\bibfnamefont{M.}~\bibnamefont{Lugaro}},
  \bibinfo{author}{\bibfnamefont{C.}~\bibnamefont{Travaglio}},
  \bibinfo{author}{\bibfnamefont{O.}~\bibnamefont{Straniero}},
  \bibinfo{author}{\bibfnamefont{A.}~\bibnamefont{Chieffi}}, \bibnamefont{and}
  \bibinfo{author}{\bibfnamefont{M.}~\bibnamefont{Limongi}},
  \bibinfo{journal}{The Astrophysical Journal} \textbf{\bibinfo{volume}{497}},
  \bibinfo{pages}{388} (\bibinfo{year}{1998}).

\bibitem[{\citenamefont{Heil et~al.}(2008)\citenamefont{Heil, Detwiler, Azuma,
  Couture, Daly, G\"orres, K\"appeler, Reifarth, Tischhauser, Ugalde
  et~al.}}]{HeilPRC2008}
\bibinfo{author}{\bibfnamefont{M.}~\bibnamefont{Heil}},
  \bibinfo{author}{\bibfnamefont{R.}~\bibnamefont{Detwiler}},
  \bibinfo{author}{\bibfnamefont{R.~E.} \bibnamefont{Azuma}},
  \bibinfo{author}{\bibfnamefont{A.}~\bibnamefont{Couture}},
  \bibinfo{author}{\bibfnamefont{J.}~\bibnamefont{Daly}},
  \bibinfo{author}{\bibfnamefont{J.}~\bibnamefont{G\"orres}},
  \bibinfo{author}{\bibfnamefont{F.}~\bibnamefont{K\"appeler}},
  \bibinfo{author}{\bibfnamefont{R.}~\bibnamefont{Reifarth}},
  \bibinfo{author}{\bibfnamefont{P.}~\bibnamefont{Tischhauser}},
  \bibinfo{author}{\bibfnamefont{C.}~\bibnamefont{Ugalde}},
  \bibnamefont{et~al.}, \bibinfo{journal}{Phys. Rev. C}
  \textbf{\bibinfo{volume}{78}}, \bibinfo{pages}{025803}
  (\bibinfo{year}{2008}).

\bibitem[{\citenamefont{Trippella et~al.}(2014)\citenamefont{Trippella, Busso,
  Maiorca, K{\"a}ppeler, and Palmerini}}]{TrippellaAPJ2014}
\bibinfo{author}{\bibfnamefont{O.}~\bibnamefont{Trippella}},
  \bibinfo{author}{\bibfnamefont{M.}~\bibnamefont{Busso}},
  \bibinfo{author}{\bibfnamefont{E.}~\bibnamefont{Maiorca}},
  \bibinfo{author}{\bibfnamefont{F.}~\bibnamefont{K{\"a}ppeler}},
  \bibnamefont{and}
  \bibinfo{author}{\bibfnamefont{S.}~\bibnamefont{Palmerini}},
  \bibinfo{journal}{The Astrophysical Journal} \textbf{\bibinfo{volume}{787}},
  \bibinfo{pages}{41} (\bibinfo{year}{2014}).

\bibitem[{\citenamefont{Aliotta et~al.}(2016)\citenamefont{Aliotta, Junker,
  Prati, Straniero, and Strieder}}]{AliottaEPJA2016}
\bibinfo{author}{\bibfnamefont{M.}~\bibnamefont{Aliotta}},
  \bibinfo{author}{\bibfnamefont{M.}~\bibnamefont{Junker}},
  \bibinfo{author}{\bibfnamefont{P.}~\bibnamefont{Prati}},
  \bibinfo{author}{\bibfnamefont{O.}~\bibnamefont{Straniero}},
  \bibnamefont{and} \bibinfo{author}{\bibfnamefont{F.}~\bibnamefont{Strieder}},
  \bibinfo{journal}{European Physical Journal A} \textbf{\bibinfo{volume}{52}},
  \bibinfo{pages}{76} (\bibinfo{year}{2016}).

\bibitem[{\citenamefont{Cristallo et~al.}(2018)\citenamefont{Cristallo,
  La~Cognata, Massimi, Best, Palmerini, Straniero, Trippella, Busso, Ciani,
  Mingrone et~al.}}]{CristalloAPJ2018}
\bibinfo{author}{\bibfnamefont{S.}~\bibnamefont{Cristallo}},
  \bibinfo{author}{\bibfnamefont{M.}~\bibnamefont{La~Cognata}},
  \bibinfo{author}{\bibfnamefont{C.}~\bibnamefont{Massimi}},
  \bibinfo{author}{\bibfnamefont{A.}~\bibnamefont{Best}},
  \bibinfo{author}{\bibfnamefont{S.}~\bibnamefont{Palmerini}},
  \bibinfo{author}{\bibfnamefont{O.}~\bibnamefont{Straniero}},
  \bibinfo{author}{\bibfnamefont{O.}~\bibnamefont{Trippella}},
  \bibinfo{author}{\bibfnamefont{M.}~\bibnamefont{Busso}},
  \bibinfo{author}{\bibfnamefont{G.~F.} \bibnamefont{Ciani}},
  \bibinfo{author}{\bibfnamefont{F.}~\bibnamefont{Mingrone}},
  \bibnamefont{et~al.}, \bibinfo{journal}{The Astrophysical Journal}
  \textbf{\bibinfo{volume}{859}}, \bibinfo{pages}{105} (\bibinfo{year}{2018}).

\bibitem[{\citenamefont{Feynman}(1972)}]{FeynmanBook1972}
\bibinfo{author}{\bibfnamefont{R.~P.} \bibnamefont{Feynman}},
  \emph{\bibinfo{title}{Statistical Mechanics, A Set of Lectures}}
  (\bibinfo{publisher}{Institute of Technology}, \bibinfo{address}{California},
  \bibinfo{year}{1972}).

\bibitem[{\citenamefont{Itoh}(1981)}]{ItohPTPS1981}
\bibinfo{author}{\bibfnamefont{N.}~\bibnamefont{Itoh}},
  \bibinfo{journal}{Progress of Theoretical Physics Supplement}
  \textbf{\bibinfo{volume}{70}}, \bibinfo{pages}{132} (\bibinfo{year}{1981}).

\bibitem[{\citenamefont{Itoh et~al.}(1977)\citenamefont{Itoh, Totsuji, and
  Ichimaru}}]{ItohApJ1977}
\bibinfo{author}{\bibfnamefont{N.}~\bibnamefont{Itoh}},
  \bibinfo{author}{\bibfnamefont{H.}~\bibnamefont{Totsuji}}, \bibnamefont{and}
  \bibinfo{author}{\bibfnamefont{S.}~\bibnamefont{Ichimaru}},
  \bibinfo{journal}{The Astrophysical Journal} \textbf{\bibinfo{volume}{218}},
  \bibinfo{pages}{477} (\bibinfo{year}{1977}).

\bibitem[{\citenamefont{Itoh et~al.}(1979)\citenamefont{Itoh, Totsuji,
  Ichimaru, and Dewitt}}]{ItohApJ1979}
\bibinfo{author}{\bibfnamefont{N.}~\bibnamefont{Itoh}},
  \bibinfo{author}{\bibfnamefont{H.}~\bibnamefont{Totsuji}},
  \bibinfo{author}{\bibfnamefont{S.}~\bibnamefont{Ichimaru}}, \bibnamefont{and}
  \bibinfo{author}{\bibfnamefont{H.~E.} \bibnamefont{Dewitt}},
  \bibinfo{journal}{The Astrophysical Journal} \textbf{\bibinfo{volume}{234}},
  \bibinfo{pages}{1079} (\bibinfo{year}{1979}).

\bibitem[{\citenamefont{Crank and Nicolson}(1947)}]{CrankMPCPS1947}
\bibinfo{author}{\bibfnamefont{J.}~\bibnamefont{Crank}} \bibnamefont{and}
  \bibinfo{author}{\bibfnamefont{P.}~\bibnamefont{Nicolson}},
  \bibinfo{journal}{Mathematical Proceedings of the Cambridge Philosophical
  Society} \textbf{\bibinfo{volume}{43}}, \bibinfo{pages}{50}
  (\bibinfo{year}{1947}).

\bibitem[{\citenamefont{Hairer et~al.}(1993)\citenamefont{Hairer, N{\o}rsett,
  and Wanner}}]{HairerBook1993}
\bibinfo{author}{\bibfnamefont{E.}~\bibnamefont{Hairer}},
  \bibinfo{author}{\bibfnamefont{S.~P.} \bibnamefont{N{\o}rsett}},
  \bibnamefont{and} \bibinfo{author}{\bibfnamefont{G.}~\bibnamefont{Wanner}},
  \emph{\bibinfo{title}{Solving Ordinary Differential Equations I}}
  (\bibinfo{publisher}{Springer-Verlag}, \bibinfo{address}{Berlin Heidelberg},
  \bibinfo{year}{1993}).

\bibitem[{\citenamefont{Schmid and Veisz}(2012)}]{SchmidRSI2012}
\bibinfo{author}{\bibfnamefont{K.}~\bibnamefont{Schmid}} \bibnamefont{and}
  \bibinfo{author}{\bibfnamefont{L.}~\bibnamefont{Veisz}},
  \bibinfo{journal}{Review of Scientific Instruments}
  \textbf{\bibinfo{volume}{83}}, \bibinfo{pages}{053304}
  (\bibinfo{year}{2012}).

\bibitem[{\citenamefont{Sylla et~al.}(2012)\citenamefont{Sylla, Veltcheva,
  Kahaly, Flacco, and Malka}}]{SyllaRSI2012}
\bibinfo{author}{\bibfnamefont{F.}~\bibnamefont{Sylla}},
  \bibinfo{author}{\bibfnamefont{M.}~\bibnamefont{Veltcheva}},
  \bibinfo{author}{\bibfnamefont{S.}~\bibnamefont{Kahaly}},
  \bibinfo{author}{\bibfnamefont{A.}~\bibnamefont{Flacco}}, \bibnamefont{and}
  \bibinfo{author}{\bibfnamefont{V.}~\bibnamefont{Malka}},
  \bibinfo{journal}{Review of Scientific Instruments}
  \textbf{\bibinfo{volume}{83}}, \bibinfo{pages}{033507}
  (\bibinfo{year}{2012}).

\bibitem[{\citenamefont{Saemann et~al.}(1999)\citenamefont{Saemann, Eidmann,
  Golovkin, Mancini, Andersson, F\"orster, and Witte}}]{Saemann1999PRL}
\bibinfo{author}{\bibfnamefont{A.}~\bibnamefont{Saemann}},
  \bibinfo{author}{\bibfnamefont{K.}~\bibnamefont{Eidmann}},
  \bibinfo{author}{\bibfnamefont{I.~E.} \bibnamefont{Golovkin}},
  \bibinfo{author}{\bibfnamefont{R.~C.} \bibnamefont{Mancini}},
  \bibinfo{author}{\bibfnamefont{E.}~\bibnamefont{Andersson}},
  \bibinfo{author}{\bibfnamefont{E.}~\bibnamefont{F\"orster}},
  \bibnamefont{and} \bibinfo{author}{\bibfnamefont{K.}~\bibnamefont{Witte}},
  \bibinfo{journal}{Phys. Rev. Lett.} \textbf{\bibinfo{volume}{82}},
  \bibinfo{pages}{4843} (\bibinfo{year}{1999}).

\bibitem[{\citenamefont{Audebert et~al.}(2002)\citenamefont{Audebert, Shepherd,
  Fournier, Peyrusse, Price, Lee, Springer, Gauthier, and
  Klein}}]{Audebert2002PRL}
\bibinfo{author}{\bibfnamefont{P.}~\bibnamefont{Audebert}},
  \bibinfo{author}{\bibfnamefont{R.}~\bibnamefont{Shepherd}},
  \bibinfo{author}{\bibfnamefont{K.~B.} \bibnamefont{Fournier}},
  \bibinfo{author}{\bibfnamefont{O.}~\bibnamefont{Peyrusse}},
  \bibinfo{author}{\bibfnamefont{D.}~\bibnamefont{Price}},
  \bibinfo{author}{\bibfnamefont{R.}~\bibnamefont{Lee}},
  \bibinfo{author}{\bibfnamefont{P.}~\bibnamefont{Springer}},
  \bibinfo{author}{\bibfnamefont{J.-C.} \bibnamefont{Gauthier}},
  \bibnamefont{and} \bibinfo{author}{\bibfnamefont{L.}~\bibnamefont{Klein}},
  \bibinfo{journal}{Phys. Rev. Lett.} \textbf{\bibinfo{volume}{89}},
  \bibinfo{pages}{265001} (\bibinfo{year}{2002}).

\bibitem[{\citenamefont{Sentoku et~al.}(2007)\citenamefont{Sentoku, Kemp,
  Presura, Bakeman, and Cowan}}]{Sentoku2007POP}
\bibinfo{author}{\bibfnamefont{Y.}~\bibnamefont{Sentoku}},
  \bibinfo{author}{\bibfnamefont{A.~J.} \bibnamefont{Kemp}},
  \bibinfo{author}{\bibfnamefont{R.}~\bibnamefont{Presura}},
  \bibinfo{author}{\bibfnamefont{M.~S.} \bibnamefont{Bakeman}},
  \bibnamefont{and} \bibinfo{author}{\bibfnamefont{T.~E.} \bibnamefont{Cowan}},
  \bibinfo{journal}{Physics of Plasmas} \textbf{\bibinfo{volume}{14}},
  \bibinfo{pages}{122701} (\bibinfo{year}{2007}).

\bibitem[{\citenamefont{Wu et~al.}(2018)\citenamefont{Wu, Gunst, Keitel, and
  P\'alffy}}]{Wu2018PRL}
\bibinfo{author}{\bibfnamefont{Y.}~\bibnamefont{Wu}},
  \bibinfo{author}{\bibfnamefont{J.}~\bibnamefont{Gunst}},
  \bibinfo{author}{\bibfnamefont{C.~H.} \bibnamefont{Keitel}},
  \bibnamefont{and} \bibinfo{author}{\bibfnamefont{A.}~\bibnamefont{P\'alffy}},
  \bibinfo{journal}{Phys. Rev. Lett.} \textbf{\bibinfo{volume}{120}},
  \bibinfo{pages}{052504} (\bibinfo{year}{2018}).

\bibitem[{\citenamefont{Lee et~al.}(2003)\citenamefont{Lee, Moon, Chung,
  Rozmus, Baldis, Gregori, Cauble, Landen, Wark, Ng et~al.}}]{LeeJOSAB2003}
\bibinfo{author}{\bibfnamefont{R.~W.} \bibnamefont{Lee}},
  \bibinfo{author}{\bibfnamefont{S.~J.} \bibnamefont{Moon}},
  \bibinfo{author}{\bibfnamefont{H.-K.} \bibnamefont{Chung}},
  \bibinfo{author}{\bibfnamefont{W.}~\bibnamefont{Rozmus}},
  \bibinfo{author}{\bibfnamefont{H.~A.} \bibnamefont{Baldis}},
  \bibinfo{author}{\bibfnamefont{G.}~\bibnamefont{Gregori}},
  \bibinfo{author}{\bibfnamefont{R.~C.} \bibnamefont{Cauble}},
  \bibinfo{author}{\bibfnamefont{O.~L.} \bibnamefont{Landen}},
  \bibinfo{author}{\bibfnamefont{J.~S.} \bibnamefont{Wark}},
  \bibinfo{author}{\bibfnamefont{A.}~\bibnamefont{Ng}}, \bibnamefont{et~al.},
  \bibinfo{journal}{J. Opt. Soc. Am. B} \textbf{\bibinfo{volume}{20}},
  \bibinfo{pages}{770} (\bibinfo{year}{2003}).

\bibitem[{\citenamefont{Vinko et~al.}(2012)\citenamefont{Vinko, Ciricosta, Cho,
  Engelhorn, Chung, Brown, Burian, Chalupsk{\'y}, Falcone, Graves
  et~al.}}]{VinkoNature2012}
\bibinfo{author}{\bibfnamefont{S.~M.} \bibnamefont{Vinko}},
  \bibinfo{author}{\bibfnamefont{O.}~\bibnamefont{Ciricosta}},
  \bibinfo{author}{\bibfnamefont{B.~I.} \bibnamefont{Cho}},
  \bibinfo{author}{\bibfnamefont{K.}~\bibnamefont{Engelhorn}},
  \bibinfo{author}{\bibfnamefont{H.~K.} \bibnamefont{Chung}},
  \bibinfo{author}{\bibfnamefont{C.~R.~D.} \bibnamefont{Brown}},
  \bibinfo{author}{\bibfnamefont{T.}~\bibnamefont{Burian}},
  \bibinfo{author}{\bibfnamefont{J.}~\bibnamefont{Chalupsk{\'y}}},
  \bibinfo{author}{\bibfnamefont{R.~W.} \bibnamefont{Falcone}},
  \bibinfo{author}{\bibfnamefont{C.}~\bibnamefont{Graves}},
  \bibnamefont{et~al.}, \bibinfo{journal}{Nature}
  \textbf{\bibinfo{volume}{482}}, \bibinfo{pages}{59} (\bibinfo{year}{2012}).

\bibitem[{\citenamefont{Hayes et~al.}(2020)\citenamefont{Hayes, Gooden, Henry,
  Jungman, Wilhelmy, Rundberg, Yeamans, Kyrala, Cerjan, Danielson
  et~al.}}]{HayesNP2020}
\bibinfo{author}{\bibfnamefont{A.~C.} \bibnamefont{Hayes}},
  \bibinfo{author}{\bibfnamefont{M.~E.} \bibnamefont{Gooden}},
  \bibinfo{author}{\bibfnamefont{E.}~\bibnamefont{Henry}},
  \bibinfo{author}{\bibfnamefont{G.}~\bibnamefont{Jungman}},
  \bibinfo{author}{\bibfnamefont{J.~B.} \bibnamefont{Wilhelmy}},
  \bibinfo{author}{\bibfnamefont{R.~S.} \bibnamefont{Rundberg}},
  \bibinfo{author}{\bibfnamefont{C.}~\bibnamefont{Yeamans}},
  \bibinfo{author}{\bibfnamefont{G.}~\bibnamefont{Kyrala}},
  \bibinfo{author}{\bibfnamefont{C.}~\bibnamefont{Cerjan}},
  \bibinfo{author}{\bibfnamefont{D.~L.} \bibnamefont{Danielson}},
  \bibnamefont{et~al.}, \bibinfo{journal}{Nature Physics}
  \textbf{\bibinfo{volume}{16}}, \bibinfo{pages}{432} (\bibinfo{year}{2020}).

\bibitem[{\citenamefont{Xu et~al.}(2013)\citenamefont{Xu, Takahashi, Goriely,
  Arnould, Ohta, and Utsunomiya}}]{XuNPA2013}
\bibinfo{author}{\bibfnamefont{Y.}~\bibnamefont{Xu}},
  \bibinfo{author}{\bibfnamefont{K.}~\bibnamefont{Takahashi}},
  \bibinfo{author}{\bibfnamefont{S.}~\bibnamefont{Goriely}},
  \bibinfo{author}{\bibfnamefont{M.}~\bibnamefont{Arnould}},
  \bibinfo{author}{\bibfnamefont{M.}~\bibnamefont{Ohta}}, \bibnamefont{and}
  \bibinfo{author}{\bibfnamefont{H.}~\bibnamefont{Utsunomiya}},
  \bibinfo{journal}{Nuclear Physics A} \textbf{\bibinfo{volume}{918}},
  \bibinfo{pages}{61} (\bibinfo{year}{2013}).

\bibitem[{\citenamefont{Dong et~al.}(2017)\citenamefont{Dong, Oganov,
  Goncharov, Stavrou, Lobanov, Saleh, Qian, Zhu, Gatti, Deringer
  et~al.}}]{DongNC2017}
\bibinfo{author}{\bibfnamefont{X.}~\bibnamefont{Dong}},
  \bibinfo{author}{\bibfnamefont{A.~R.} \bibnamefont{Oganov}},
  \bibinfo{author}{\bibfnamefont{A.~F.} \bibnamefont{Goncharov}},
  \bibinfo{author}{\bibfnamefont{E.}~\bibnamefont{Stavrou}},
  \bibinfo{author}{\bibfnamefont{S.}~\bibnamefont{Lobanov}},
  \bibinfo{author}{\bibfnamefont{G.}~\bibnamefont{Saleh}},
  \bibinfo{author}{\bibfnamefont{G.-R.} \bibnamefont{Qian}},
  \bibinfo{author}{\bibfnamefont{Q.}~\bibnamefont{Zhu}},
  \bibinfo{author}{\bibfnamefont{C.}~\bibnamefont{Gatti}},
  \bibinfo{author}{\bibfnamefont{V.~L.} \bibnamefont{Deringer}},
  \bibnamefont{et~al.}, \bibinfo{journal}{Nature Chemistry}
  \textbf{\bibinfo{volume}{9}}, \bibinfo{pages}{440} (\bibinfo{year}{2017}).

\bibitem[{\citenamefont{Harissopulos et~al.}(2005)\citenamefont{Harissopulos,
  Becker, Hammer, Lagoyannis, Rolfs, and Strieder}}]{HarissopulosPRC2005}
\bibinfo{author}{\bibfnamefont{S.}~\bibnamefont{Harissopulos}},
  \bibinfo{author}{\bibfnamefont{H.~W.} \bibnamefont{Becker}},
  \bibinfo{author}{\bibfnamefont{J.~W.} \bibnamefont{Hammer}},
  \bibinfo{author}{\bibfnamefont{A.}~\bibnamefont{Lagoyannis}},
  \bibinfo{author}{\bibfnamefont{C.}~\bibnamefont{Rolfs}}, \bibnamefont{and}
  \bibinfo{author}{\bibfnamefont{F.}~\bibnamefont{Strieder}},
  \bibinfo{journal}{Phys. Rev. C} \textbf{\bibinfo{volume}{72}},
  \bibinfo{pages}{062801(R)} (\bibinfo{year}{2005}).

\bibitem[{\citenamefont{{Helmholtz International Beamline for Extreme Fields at
  the European XFEL}}(2022)}]{HIBEF-web}
\bibinfo{author}{\bibnamefont{{Helmholtz International Beamline for Extreme
  Fields at the European XFEL}}}, \bibinfo{howpublished}{Official Website}
  (\bibinfo{year}{2022}), \bibinfo{note}{{
  {https://www.hzdr.de/db/Cms?pNid=694}}}.

\bibitem[{\citenamefont{Wu et~al.}(2019)\citenamefont{Wu, Keitel, and
  P\'alffy}}]{WuPRA2019}
\bibinfo{author}{\bibfnamefont{Y.}~\bibnamefont{Wu}},
  \bibinfo{author}{\bibfnamefont{C.~H.} \bibnamefont{Keitel}},
  \bibnamefont{and} \bibinfo{author}{\bibfnamefont{A.}~\bibnamefont{P\'alffy}},
  \bibinfo{journal}{Phys. Rev. A} \textbf{\bibinfo{volume}{100}},
  \bibinfo{pages}{063420} (\bibinfo{year}{2019}).

\bibitem[{\citenamefont{Berlinguette et~al.}(2019)\citenamefont{Berlinguette,
  Chiang, Munday, Schenkel, Fork, Koningstein, and
  Trevithick}}]{BerlinguetteNature2019}
\bibinfo{author}{\bibfnamefont{C.~P.} \bibnamefont{Berlinguette}},
  \bibinfo{author}{\bibfnamefont{Y.-M.} \bibnamefont{Chiang}},
  \bibinfo{author}{\bibfnamefont{J.~N.} \bibnamefont{Munday}},
  \bibinfo{author}{\bibfnamefont{T.}~\bibnamefont{Schenkel}},
  \bibinfo{author}{\bibfnamefont{D.~K.} \bibnamefont{Fork}},
  \bibinfo{author}{\bibfnamefont{R.}~\bibnamefont{Koningstein}},
  \bibnamefont{and} \bibinfo{author}{\bibfnamefont{M.~D.}
  \bibnamefont{Trevithick}}, \bibinfo{journal}{Nature}
  \textbf{\bibinfo{volume}{570}}, \bibinfo{pages}{45} (\bibinfo{year}{2019}).

\end{thebibliography}

\end{document}